\documentclass[12pt,preprint]{aastex}

\shorttitle{X-ray Observations of Dwarf Novae}
\shortauthors{Pandel et al.}

\newcommand{\xmm}{{\it XMM-Newton}}
\newcommand{\chandra}{{\it Chandra}}
\newcommand{\xspec}{{XSPEC}}
\newcommand{\mekal}{{MEKAL}}
\newcommand{\mkcflow}{{MKCFLOW}}
\newcommand{\pcfabs}{{PCFABS}}
\newcommand{\hrefl}{{HREFL}}
\newcommand{\tnm}[1]{\makebox[1em][l]{$^\mathrm{#1}$}\hspace{-1em}}
\newcommand{\tnt}[2]{\hspace{0.1em}\makebox[0.5em][l]{$^\mathrm{{#1},{#2}}$}}
\newcommand{\dpm}[3]{${#1}^{+{#2}}_{-{#3}}$}

\begin{document}


\title{X-ray Observations of the Boundary Layer in Dwarf Novae at Low Accretion Rates}

\author{Dirk Pandel}
\affil{Department of Physics, University of California, Santa Barbara, CA 93106, USA}
\email{dpandel@xmmom.physics.ucsb.edu}
\and
\author{France A. C\'ordova}
\affil{Institute of Geophysics and Planetary Physics, Department of Physics,\\
University of California, Riverside, CA 92521, USA}
\and
\author{Keith O. Mason}
\affil{Mullard Space Science Laboratory, Department of Space and Climate Physics,\\
University College London, Holmbury St. Mary, Dorking, Surrey RH5 6NT, UK}
\and
\author{William C. Priedhorsky}
\affil{Los Alamos National Laboratory, International, Space, and Response Division,\\
MS B241, Los Alamos, NM 87545, USA}


\begin{abstract}

We present a spectral analysis of \xmm\ data of ten dwarf novae, nine of which were observed
during a quiescent state.
The X-ray spectra indicate the presence of a hot, optically thin plasma with a temperature
distribution consistent with a simple, isobaric cooling flow.
The likely origin of the X-ray emission is cooling plasma in the boundary layer settling onto
the white dwarf.
Using a cooling flow model, we derive the temperatures, accretion rates, rotational velocities,
and elemental abundances of the X-ray emitting gas.
We discuss the implications of our findings for the structure of the boundary layer.
A comparison of X-ray and ultraviolet luminosities finds no evidence of underluminous boundary layers
in any of the quiescent dwarf novae.
The X-ray spectrum of EI~UMa differs significantly from those of the other objects, showing a
strong fluorescent Fe K-$\alpha$ line and a strong O~{\sc vii} line triplet.
Based on the observational evidence, we argue that EI~UMa is most likely an intermediate polar and not
as previously thought a dwarf nova.

\end{abstract}

\keywords{
accretion, accretion disks ---
binaries: close ---
novae, cataclysmic variables ---
stars: dwarf novae ---
stars: individual (OY Car, WW Cet, AB Dra, U Gem, VW Hyi, WX Hyi, T Leo, TY PsA, EI UMa, SU UMa) ---
X-rays: binaries
}


\section{INTRODUCTION}

Dwarf novae belong to the class of nonmagnetic cataclysmic variables in which the magnetic field of the white dwarf
is too weak to disrupt the accretion disk, so that the disk can extend close to the surface of the white dwarf
(see \citet{1995cvs..book.....W} for a review of dwarf novae).
The material in the inner disk, initially moving with a Keplerian velocity, must dissipate its rotational
kinetic energy in order to accrete onto the slowly rotating white dwarf.
The structure of the boundary layer, i.e.\ the transition region between the disk and the white dwarf, is poorly understood,
and theoretical modelling is complicated by the strong shearing and turbulence present in the accretion flow.
A number of widely different models have been proposed, including a disk-like boundary layer \citep{1993Natur.362..820N},
a coronal siphon flow \citep{1994A&A...288..175M}, a spherical corona \citep{1999PASJ...51...45M},
and a hot settling flow \citep{2002ApJ...565L..39M}.

During their quiescent state, dwarf novae are copious sources of hard X-rays with energies of 10~keV and higher.
The origin of the X-ray emission is thought to be a hot, optically thin plasma in the boundary layer.
Observations (e.g.\ of eclipse profiles) have shown that the X-ray emitting region is generally small and
close to the white dwarf's surface \citep[e.g.][]{1996A&A...315..467V,1997ApJ...475..812M}.
Basic accretion theory predicts that half of the total accretion energy should emerge from the disk at optical and
ultraviolet (UV) wavelengths, while the other half is released in the hot boundary layer as X-ray and extreme UV
emission \citep{1974MNRAS.168..603L}.
However, in many dwarf novae the X-ray luminosity is found to be lower than predicted 
\citep[e.g.][]{1982ApJ...262L..53F,1994A&A...292..519V}.
It has been suggested that this ``mystery of the missing boundary layers'' may be due to
a disruption of the inner disk caused by irradiation from the white dwarf \citep{1997MNRAS.288L..16K},
evaporation of the inner disk via a coronal siphon flow \citep{1994A&A...288..175M},
or white dwarfs rotating near breakup velocity \citep{1995MNRAS.276..495P}.

The greatly improved sensitivity and spectral resolution of the new generation of X-ray telescopes promises
to reveal a much more detailed view of the boundary layer.
In this paper, we present a spectral analysis of \xmm\ data of ten dwarf novae at low accretion rates.
The high sensitivity of \xmm\ enables us to examine the multi-temperature nature of the boundary layer
and derive the temperature distribution of the X-ray emitting plasma.
We compare X-ray and UV luminosities in order to investigate whether boundary layer luminosity is missing
in any of the dwarf novae.
Our spectral analysis further provides estimates of the accretion rates, boundary layer rotation velocities, and
elemental abundances in the accreting material.
We use our findings to draw conclusions about the structure of the boundary layer.
Based on our X-ray spectral analysis, we argue that one of the objects, EI~UMa, is actually an intermediate polar
and not as previously thought a U~Gem type dwarf nova.
We included in our sample the dwarf nova VW~Hyi, for which we already presented a spectral and timing analysis
of the \xmm\ data in \citet{2003MNRAS.346.1231P}.
The OY~Car data has previously been presented in \citet{2001A&A...365L.288R,2001A&A...365L.294R}
and \citet{2003MNRAS.345.1009W}.
Some properties of the ten cataclysmic variables in our sample are summarized in Table~\ref{nmsyspar}.


\section{OBSERVATIONS AND DATA REDUCTION}

\begin{deluxetable}{lcccccccc}
\tablecaption{System Parameters \label{nmsyspar}}
\tablewidth{0pc}
\tabletypesize{\small}
\tablehead{
\colhead{Object} & \colhead{$P_{orb}$} & \colhead{$D$} &
\colhead{Type} & \colhead{$T_{ob}$} & \colhead{$T_{sob}$} &
\colhead{$i$} & \colhead{$M_{wd}$} & \colhead{$R_{wd}$} \\
\colhead{Name} & \colhead{(min)} & \colhead{(pc)} &
\colhead{} & \colhead{(d)} & \colhead{(d)} &
\colhead{(deg)} & \colhead{($M_\odot$)} & \colhead{($10^8$cm)}}
\startdata
T Leo  & \phn84.7 & $92\pm15$\tnm{e}        & SU        & ?          & 420 & $65\pm19$        & 0.4\tnm{d}      &    10.5 \\
OY Car & \phn90.9 & $86\pm4$\tnm{a}\phn     & SU        & 14         & 325 & $83.3\pm0.2$\phn & $0.685\pm0.011$ & \phn7.8 \\
VW Hyi &    107.0 & 65\tnm{h}               & SU        & 28         & 183 & $60\pm10$        & $0.63\pm0.15$   & \phn8.3 \\
WX Hyi &    107.7 & 265\tnm{h}              & SU        & 11         & 180 & $40\pm10$        & $0.90\pm0.30$   & \phn6.2 \\
SU UMa &    109.9 & 280\tnm{h}              & SU        & \phn5--33  & 160 & 44\tnm{h}        & 0.7\tnm{*}      & \phn7.7 \\
TY PsA &    121.1 & 190\tnm{h}              & SU        & 30--50     & 202 & 65\tnm{h}        & 0.7\tnm{*}      & \phn7.7 \\
AB Dra &    218.9 & 90\tnm{g}               & ZC        & \phn8--22  & --  & $40\pm20$\tnm{c} & 0.7\tnm{*}      & \phn7.7 \\
WW Cet &    253.2 & $146\pm25$\tnm{e}\phn   & ZC\tnm{h} & 45         & --  & $54\pm4$\phn     & $0.85\pm0.11$   & \phn6.5 \\
U Gem  &    254.7 & $96.4\pm4.6$\tnm{b}\phn & UG        & 132        & --  & $69\pm2$\phn     & $1.07\pm0.08$   & \phn4.9 \\
EI UMa &    386.1 & 100\tnm{*}              & \phn UG?  & ?          & --  & 23\tnm{f}        & 0.95\tnm{*}     & \phn5.9
\enddata
\tablecomments{
The system parameters shown are the
orbital period $P_{orb}$, distance $D$, dwarf nova type (U~Gem, SU~UMa, or Z~Cam),
average time between outbursts $T_{ob}$ and superoutbursts $T_{sob}$,
orbital inclination $i$, white dwarf mass $M_{wd}$, and white dwarf radius $R_{wd}$.
Unless noted otherwise, the parameters are from \citet{2003A&A...404..301R}.
Parameter values followed by $^*$ are not known and have been assumed (for $M_{wd}$ of EI~UMa see Section~\ref{eiumadisc}).
$R_{wd}$ was derived from $M_{wd}$ using the mass-radius relationship in \citet{1961ApJ...134..683H}.
}
\tablenotetext{a}{\citet{1996A&A...306..151B}}
\tablenotetext{b}{\citet{1999ApJ...515L..93H}}
\tablenotetext{c}{\citet{1991A&A...252..100L}}
\tablenotetext{d}{\citet{1984ApJ...276..305S}}
\tablenotetext{e}{\citet{1996MNRAS.282.1211S}}
\tablenotetext{f}{\citet{1986AJ.....91..940T}}
\tablenotetext{g}{\citet{1982AJ.....87.1558W}}
\tablenotetext{h}{\citet{1987MNRAS.227...23W}}
\end{deluxetable}

\begin{deluxetable}{lccccccccccc}
\tablecaption{Summary of \xmm\ Observations \label{nmobspar}}
\tablewidth{0pc}
\rotate
\tabletypesize{\small}
\tablehead{
Object & Start & \multicolumn{4}{c}{Effective Exposure Times} & \multicolumn{3}{c}{X-ray Count Rates (s$^{-1}$)} & UV Flux & B Mag- & State \\
Name   & of Obs. & \multicolumn{4}{c}{(ks)} & \multicolumn{3}{c}{(0.2--12~keV)} & at 290~nm & nitude &  \\
       & (MJD) & MOS & PN & RGS & OM & MOS-1 & MOS-2 & PN & (mJy) &  &  }
\startdata
T Leo  & 52426.363 &    12.3        &    10.0        &    12.8        & \phn7.8 & 1.09\tnt{l}{m} & 1.18\tnt{s}{t} & 3.86\tnt{f}{m} & 1.65         & 16.2        & Q \phn\phn \\
OY Car & 51724.939 &    51.4\tnm{*} &    48.9\tnm{*} &    52.2\tnm{*} & \phn8.4 & 0.39\tnt{f}{m} & 0.39\tnt{f}{m} & 1.19\tnt{f}{m} & 0.68/0.94    & --          & Q \phn4    \\
       & 51763.327 &    13.7\tnm{*} & \phn6.2\tnm{*} &    13.0\tnm{*} &    10.0 & 0.20\tnt{f}{m} & 0.20\tnt{f}{m} & 0.59\tnt{f}{m} & --           & 16.8/16.2   & Q 42       \\
VW Hyi & 52201.222 &    18.7        &    16.1        &    19.3        &    17.3 & 0.67\tnt{s}{m} & 0.71\tnt{s}{t} & 2.41\tnt{f}{m} & \phn7.5/10.6 & 14.4/13.9   & Q 22       \\
WX Hyi & 52282.148 & \phn9.8\tnm{*} & \phn7.5\tnm{*} &    10.3\tnm{*} &    12.9 & 0.24\tnt{l}{m} & 0.26\tnt{s}{t} & 0.87\tnt{f}{m} & 17.7         & 13.1        & O \phn\phn \\
SU UMa & 52399.717 &    13.8\tnm{*} & \phn9.1\tnm{*} &    11.1\tnm{*} & \phn9.0 & 2.19\tnt{s}{m} & 2.31\tnt{s}{t} & 6.95\tnt{f}{m} & 4.8          & --          & Q \phn8    \\
TY PsA & 52241.792 &    12.6        &    10.0        &    13.2        &    11.6 & 0.38\tnt{f}{m} & 0.41\tnt{s}{t} & 1.36\tnt{f}{m} & 1.00         & 16.6        & Q \phn\phn \\
AB Dra & 52553.235 &    11.7        &    10.0        & \phn9.2\tnm{*} &     --  & 0.68\tnt{f}{m} & 0.72\tnt{s}{t} & 2.40\tnt{f}{t} & --           & --          & Q \phn5    \\
WW Cet & 52249.724 &    12.1        & \phn9.5        &    12.7        &    11.2 & 1.36\tnt{l}{m} & 1.41\tnt{s}{t} & 4.50\tnt{f}{m} & 2.7          & 15.5        & Q 20       \\
U Gem  & 52377.214 &    22.4        &    21.6        &    22.9        &    20.8 & 0.93\tnt{l}{m} & 0.97\tnt{s}{t} & 3.32\tnt{s}{m} & 5.2          & 14.5        & Q 50       \\
EI UMa & 52404.462 & \phn7.4        & \phn4.5\tnm{*} & \phn6.6\tnm{*} & \phn7.5 & 1.93\tnt{l}{m} & 2.08\tnt{s}{t} & 6.64\tnt{f}{m} & 4.1          & 14.8        & Q \phn\phn \\
       & 52580.859 & \phn9.9        & \phn8.2        &    10.0        & \phn8.5 & 2.03\tnt{l}{m} & 2.25\tnt{s}{t} & 6.89\tnt{f}{m} & 5.9          & 14.2        & I \phn\phn 
\enddata
\tablecomments{
The table summarizes \xmm\ observing dates, effective exposure times for all instruments,
average EPIC count rates, as well as UV fluxes (UVW1-filter) and B magnitudes obtained with the OM.
A $^*$ following the exposure times indicates that some part of the exposure has been excluded
due to high background.
The superscripts following the EPIC count rates indicate the window mode (f~-~full frame, l~-~large, s~-~small)
and the filter (t~-~thin, m~-~medium).
The OM light curves for OY~Car and VW~Hyi showed significant orbital variations,
so we quote the low and high brightness levels (not considering the eclipse in OY~Car).
The last column shows the state of the systems (Q~-~quiescent, I~-~intermediate, O~-~outburst) as 
derived from AAVSO light curves and B magnitudes
followed by the number of days since the end of the last outburst (when known).
}
\end{deluxetable}

We present X-ray and UV data obtained with \xmm\ \citep{2001A&A...365L...1J}
during twelve observations of ten dwarf novae.
Our analysis includes data from the
EPIC MOS and PN cameras \citep{2001A&A...365L..27T,2001A&A...365L..18S},
the RGS \citep{2001A&A...365L...7D},
and the Optical Monitor \citep{2001A&A...365L..36M}.
A summary of the observations is given in Table~\ref{nmobspar}.
We derived the state of the binaries from light curves provided by the AAVSO
and from the B magnitudes measured with the \xmm\ Optical Monitor (OM).
All but two of the observations occurred during quiescence.
When known, we quote in the last column of Table~\ref{nmobspar} the number of days
since the end of the last outburst.
All of the SU~UMa type dwarf novae were observed after a normal outburst and
none after a superoutburst.

From the EPIC MOS and PN data, we extracted source photons using a circular 
aperture with a radius of $40^{\prime\prime}$.
We included in our analysis good photon events ($\mathrm{FLAG}=0$) with patterns 0--12 for 
MOS and 0--4 for PN.
Background rates were estimated using larger regions on the same CCD as the source image
(annular regions for MOS and off-center regions for PN)
unless the EPIC cameras were operated in small window mode.
In this case, regions on other CCDs had to be used.
Vignetting corrections appropriate for the selected regions were applied to all background rates.
For some observations we excluded time periods with high background
(indicated by a $^*$ following the exposure times in Table~\ref{nmobspar}).
After this exclusion, average EPIC background rates in the energy range 0.2--12~keV
were typically $\sim$1\% of the source count rate with the exception of AB~Dra
(7\%) and WX~Hyi (5\%).
For our analysis of the RGS data, we used the first order spectra only.

We derived the UV fluxes and B magnitudes in Table~\ref{nmobspar} from the data obtained
with the Optical Monitor.
The OM was operated in fast mode which, in addition to images, provided individual photon
arrival times inside a small window with a time resolution of 0.5~s.
With the exception of the OY~Car observations, the OM was configured to perform one 1.2~ks exposure
with the B filter followed by several longer exposures with the UVW1 filter (240--340~nm).
Due to technical problems, the data for one or both filters were lost during the AB~Dra and SU~UMa observations.
No fast mode data were available for the U~Gem observation, so we derived the B/UV brightness
from the OM images.
We extracted source photons using a circular aperture with a $5^{\prime\prime}$ radius and corrected
the count rates for the 81\% enclosed energy fraction of this aperture.
We adopted for a count rate of 1~s$^{-1}$ in the UVW1 filter a flux conversion of 0.126~mJy
($4.5\times10^{-16}$~erg~cm$^{-2}$~s$^{-1}$~\AA$^{-1}$) at 290~nm, which is appropriate for
UV-bright objects\footnote{http://xmm.vilspa.esa.es/sas/documentation/watchout/uvflux.shtml}.
In the B filter, a count rate of 1~s$^{-1}$ corresponds to a magnitude of 19.27
({\it XMM-Newton Users' Handbook}\footnote{http://xmm.vilspa.esa.es/external/xmm\_user\_support/documentation/uhb/index.html}).
The background rates, which we estimated from the OM images, were typically 
$\sim$1~s$^{-1}$ (0.1~mJy) for the UVW1 filter and $\sim$5~s$^{-1}$ for the B filter.
We corrected the count rates for coincidence losses using the method described in the 
{\it XMM-Newton Users' Handbook}.
Coincidence losses of 10\% are expected for count rates of $\sim$50~s$^{-1}$,
corresponding to 5~mJy in the UVW1 filter and a magnitude of 15.0 in the B filter.
Due to the high count rates, the coincidence loss correction for the WX~Hyi data
is probably unreliable.
For OY~Car and VW~Hyi, which show strong orbital variations, we quote in Table~\ref{nmobspar}
the minimum and maximum brightness during the orbital cycle (excluding the eclipse in OY~Car).


\section{SPECTRAL ANALYSIS}
\label{nmspec}

\begin{figure*}
\epsscale{0.94}
\plotone{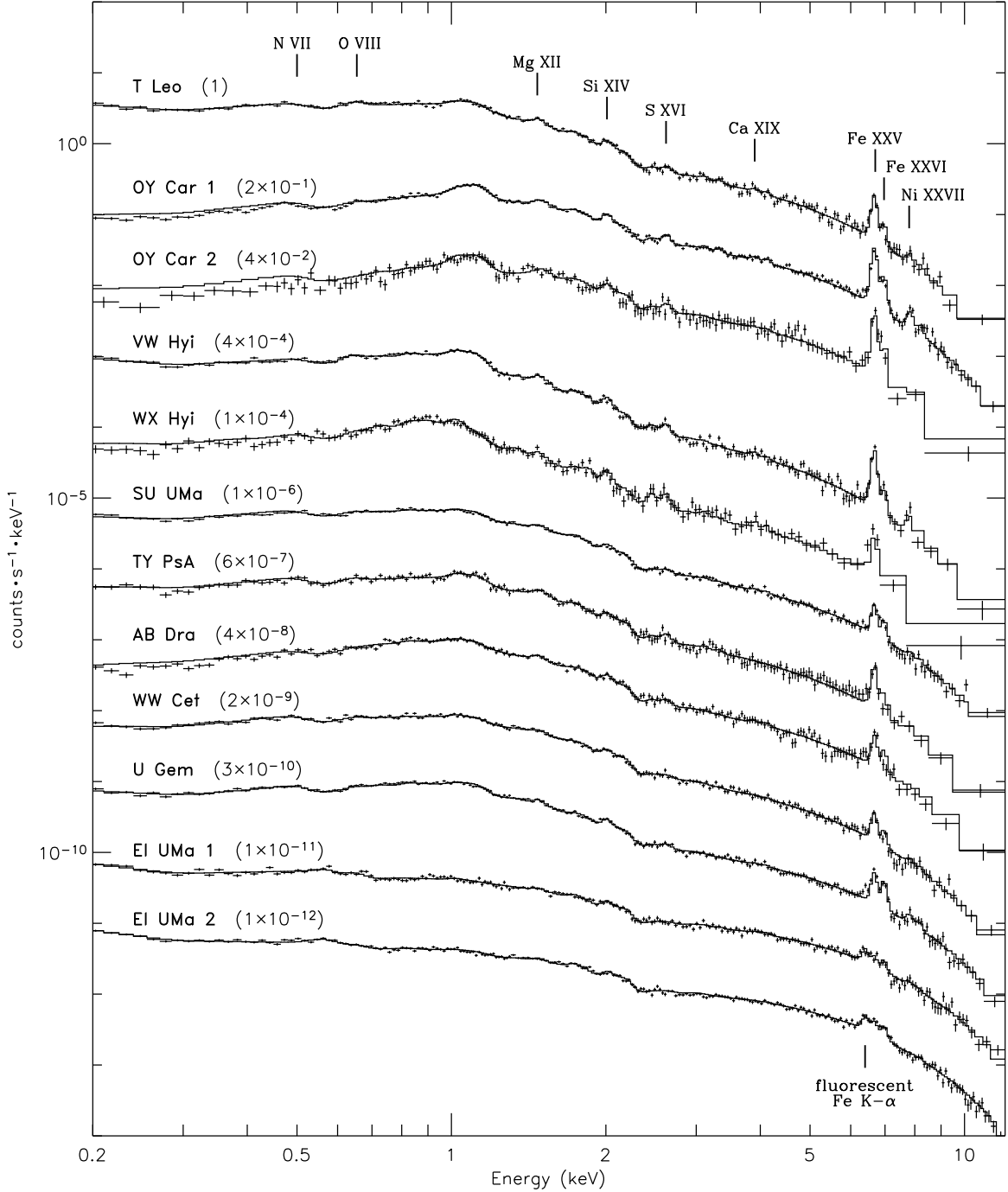}
\caption{\label{nmepic}
Combined EPIC MOS and PN spectra for the 12 observations of the 10 dwarf novae.
The solid lines show the best fits with our cooling flow model (Section~\ref{cflowmod} and Table~\ref{nmfitresults}).
For clarity, the spectra have been shifted vertically by the factors indicated in parentheses.
}
\end{figure*}

\begin{figure*}
\epsscale{0.97}
\plotone{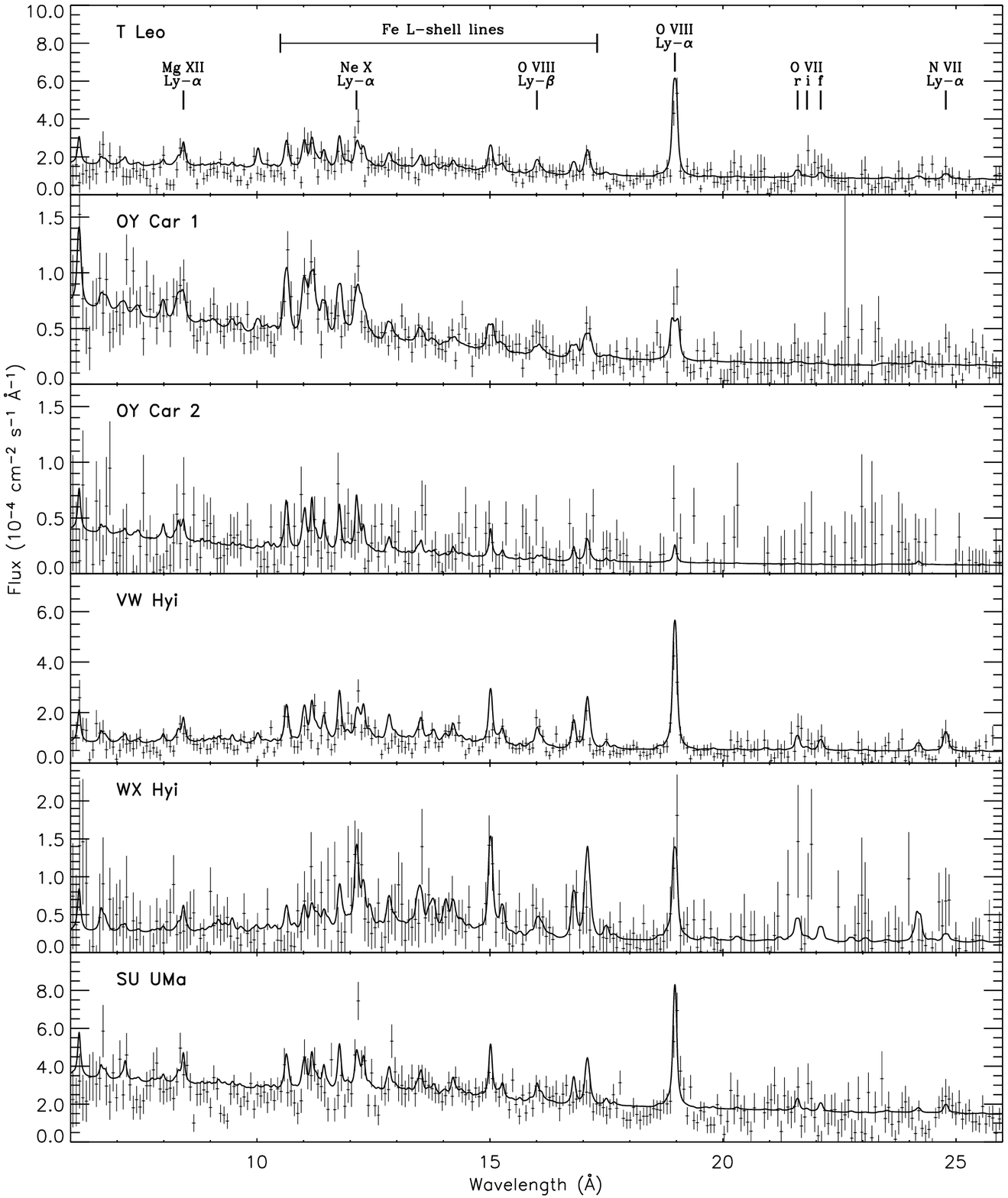}
\caption{\label{nmrgs1}
Flux calibrated RGS spectra.
The solid lines show the best fits with our cooling flow model (Section~\ref{cflowmod} and Table~\ref{nmfitresults}).
The data are binned at approximately the FWHM detector resolution (0.07~\AA).
}
\end{figure*}

\begin{figure*}
\epsscale{0.97}
\plotone{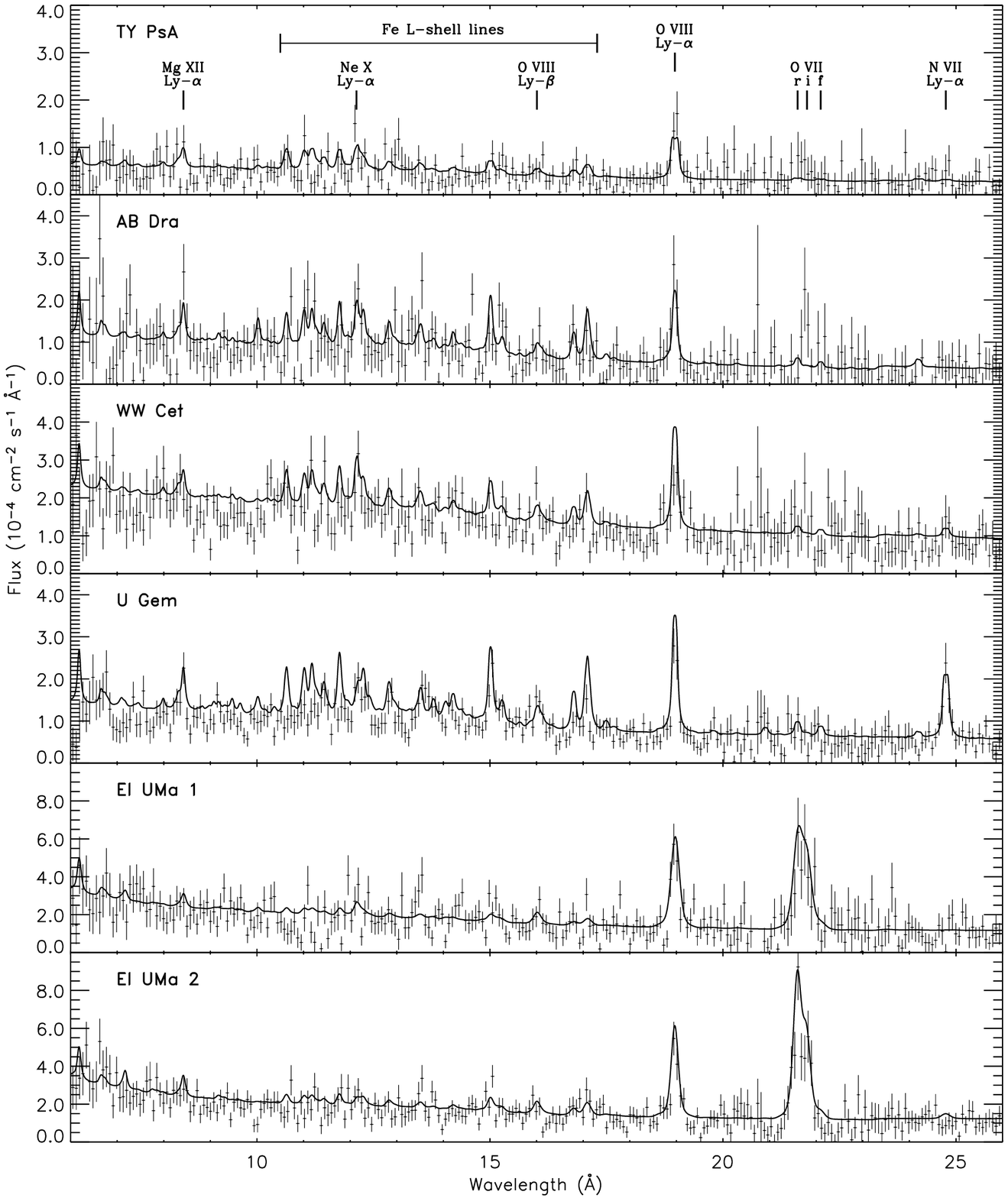}
\caption{\label{nmrgs2}
Flux calibrated RGS spectra.
The solid lines show the best fits with our cooling flow model (Section~\ref{cflowmod} and Table~\ref{nmfitresults}).
The data are binned at approximately the FWHM detector resolution (0.07~\AA).
}
\end{figure*}

The medium resolution EPIC spectra and the high resolution RGS spectra
of the ten dwarf novae are shown in Figs~\ref{nmepic}--\ref{nmrgs2}.
We performed the spectral fitting with the \xspec\ package version 11.2
\citep{1996adass...5...17A}.
For optimal fit results, we binned the EPIC and RGS spectra at one
third of the respective FWHM detector resolution.
To account for the low number of counts per bin, we used C-statistic
\citep{1979ApJ...228..939C} instead of $\chi^2$-statistic.
For clarity, the spectra in Figs~\ref{nmepic}--\ref{nmrgs2} are shown with
a lower resolution than was used for fitting.

The spectral fitting was performed simultaneously for the five spectra from
the three EPIC cameras and the two RGS spectrometers.
For the EPIC spectra we restricted the fitting to the energy range 0.2--12~keV.
We found that individual fits to the three EPIC spectra yield slightly different
results which we attribute to discrepancies in the cross-calibration.
The differences are mainly reflected in the spectral slopes (parameter $\alpha$ below),
which may differ by up to 6$\sigma$ between the three EPIC spectra
\citep[see also our analysis of the VW~Hyi data in][]{2003MNRAS.346.1231P}.
As a consequence, quality criteria for the simultaneous fits do not reflect
how well the spectral models agree with the data but rather how well the different
EPIC cameras agree with each other.
We therefore used as a quality criterion the $\chi^2$ for a fit to only the spectrum of
the EPIC PN, the detector with the highest signal-to-noise ratio.
In order to apply $\chi^2$-statistic, we rebinned the data with a minimum of 20~counts per bin.
The quoted fit parameters, however, were obtained from the combined fits to the EPIC and RGS spectra.
Because of the lower sensitivity, the RGS data have only a small influence on the fit parameters responsible
for the X-ray continuum.
However, they do complement the EPIC spectra in the determination of
elemental abundances from the emission line strengths
(see \citet{2003MNRAS.346.1231P} for a comparison between the abundances in VW~Hyi derived
from the EPIC and the RGS spectra).


\subsection{Cooling Flow Model}
\label{cflowmod}

The X-ray spectra (Figs~\ref{nmepic}--\ref{nmrgs2}) are rich in emission lines,
indicating the presence of a hot, optically thin plasma.
Most prominent are the H- and He-like emission lines of the abundant elements
from N to Fe as well as some of the Fe L-shell lines between 10 and 18~\AA\ (0.7--1.3~keV).
The presence of emission lines from many different elements is an indication
that the X-ray emitting plasma covers a wide range of temperatures.
We first attempted to fit the spectra with one or more single-temperature 
plasma models.
As the single-temperature model we used the \mekal\ model in \xspec,
which describes an optically thin and collisionally ionized, isothermal plasma
based on calculations by \citet{1985A&AS...62..197M} and \citet{1995ApJ...438L.115L}.
We found that for most spectra a one- or two-temperature model is inconsistent
with the data.
The only exceptions are the WX~Hyi spectrum and the second OY~Car spectrum.
The likely reason for the good agreement is the low signal-to-noise ratio
of both spectra.
For most of the other dwarf novae, however, the spectra are well fit 
by three- or four-temperature \mekal\ models.
Exceptions are the first OY~Car spectrum and both EI~UMa spectra, which are inconsistent
with any \mekal-based, multi-temperature model.
We will discuss these special cases further in Sections~\ref{oycar} and~\ref{eiuma}.

It is evident from the multiple plasma temperatures required in our fits
that the X-ray emitting gas has a wide range of temperatures.
This is in agreement with earlier findings for quiescent dwarf novae
\citep[e.g.][]{1996cvro.coll..259I,2002ApJ...574..942S,2003ApJ...598..545P}.
A continuous temperature distribution is expected for accreting gas that
cools as it settles onto the white dwarf.
An appropriate model to describe such a plasma is that of an isobaric cooling flow.
In a simple cooling flow model, the emission measure $EM$ at each temperature is
proportional to the time the cooling gas remains at this temperature
(i.e.\ the cooling time) and therefore inversely proportional to the bolometric
luminosity \citep[e.g.][]{1988cfcg.work...53M}.
The differential emission measure for an isobaric cooling flow is given by
\begin{equation}
\left.\frac{dEM}{dT}\right|_\mathrm{c.f.}
=\frac{5}{2}\,k
\times\frac{\dot{M}}{\mu\,m_p}
\times\frac{n^2}{\varepsilon(T,n)}\ ,
\label{emdist}
\end{equation}
where $\dot{M}$ is the accretion rate, $m_p$ the proton mass, $\mu$ the mean
molecular weight ($\sim$0.6), $n$ the particle density, and $\varepsilon(T,n)$
the total emissivity per volume (in erg~s$^{-1}$~cm$^{-3}$).
This equation follows from the assumption that the plasma is cooling isobarically,
i.e.\ the energy radiated by the plasma is $-5/2\,k\,dT$ per particle.
Furthermore, it is assumed for this model that the plasma is optically thin, that any external input of energy
(e.g.\ from the gravitational field) can be neglected, and that there is no heat conduction in the flow.
Note that for a collisionally ionized plasma in thermal equilibrium $\varepsilon(T,n)\propto n^2$,
so that the emission measure distribution is independent of the density and therefore independent of
the geometry and scale of the cooling flow.

In \xspec, a simple, isobaric cooling flow is described by the \mkcflow\ model, which assumes the above emission measure
distribution and calculates the X-ray spectrum at each temperature using the \mekal\ model.
The \mkcflow\ model agrees with our data as well as the three- and four-temperature models described above.
We find that small, but statistically significant improvements over the \mkcflow\ model can be achieved 
if the slope of the emission measure distribution is allowed to vary.
Such deviations from an isobaric cooling flow could be caused by the strong gravitational field
on the white dwarf, by the plasma becoming optically thick at lower temperatures, by heat conduction,
or by uncertainties in the detector calibration.
We modified the \mkcflow\ model so that the emission measure distribution is given by
\begin{equation}
\frac{dEM}{dT}
=\left.\frac{dEM}{dT}\right|_\mathrm{c.f.}
\times\left(\frac{T}{T_{max}}\right)^\alpha\ ,
\label{emdist2}
\end{equation}
where $T_{max}$ is the initial temperature of the cooling gas and $\alpha$ a power-law index
that parameterizes the deviation from an isobaric flow.
The other parameters in our model are the minimum temperature $T_{min}$, the accretion rate
$\dot{M}_{bl}$ at $T_{max}$ for a given distance, and the elemental abundances.
We calculated the X-ray spectra by adding single temperature \mekal\ models on a grid with a spacing
of 0.1 in $\log T$ between 8~eV and 80~keV.
To determine the neutral hydrogen column density $N_H$ of the intervening gas,
we included photoelectric absorption in our model.

The resolution of the RGS spectra is sufficient to measure the rotational velocity of the boundary layer
via Doppler broadening of emission lines.
We therefore included line broadening in our model, assuming that the X-ray emitting gas
is a thin ring rotating with a non-relativistic velocity and that half of the ring is
obscured by the white dwarf (i.e.\ the visible part is a semicircle).
The line broadening profile for an emission line at wavelength $\lambda_0$ is then given by
\begin{equation}
P(\lambda)=\pi^{-1}\left((\beta_{bl}\lambda_0)^2-(\lambda-\lambda_0)^2\right)^{-1/2}
\end{equation}
for $\left|\lambda-\lambda_0\right|<\beta_{bl}\lambda_0$ and $P(\lambda)=0$ otherwise.
Here $\beta_{bl}=V_{bl}\sin{(i)}/c$ with $V_{bl}$ being the rotation velocity, $i$ the orbital inclination,
and $c$ the speed of light.

\begin{deluxetable}{lcccccccc}
\tablecaption{Fit Parameters for the Cooling Flow Model\label{nmfitresults}}
\tablewidth{0pc}
\rotate
\tabletypesize{\small}
\tablehead{
Object &  $kT_{max}$  &  $\alpha$  &       $\dot{M}_{bl}$        &          $N_H$          &  $V_{bl}\sin{i}$  &        $L_{bl}$         &  $\chi^2$ (dof)     &  Prob.  \\
       &    (keV)     &            &  ($\times10^{-12}$\phn      &  ($\times10^{20}$\phn   & (km$\,$s$^{-1}$)  &  ($\times10^{31}$\phn   &      for PN         &         \\
       &              &            &  \phn $M_\odot\,$yr$^{-1}$) &   \phn cm$^{-2}$)       &                   &  \phn erg$\,$s$^{-1}$)  &                     &         }
\startdata
T Leo  &  $10.8\pm0.8$  &  $\phm{-}0.19\pm0.09$  &       $13\pm2$    &  $0.3\pm0.2\phn$  &  \phn\dpm{710}{240}{380}  &   3.0      &     204 (180)  &  11\%     \\[0.5ex]
OY Car &  $11.7\pm0.9$  &  $\phm{-}0.35\pm0.07$  &  $\phn6.8\pm0.7$  &  $2.8\pm0.3\phn$  &     \dpm{1350}{400}{400}  &   1.5      &     305 (211)  &  \phn0\%  \\[0.5ex]
       &  $11.0\pm0.3$  &  $\phm{-}0.26\pm0.07$  &  $\phn4.0\pm0.1$  &  $2.9\pm0.6\phn$  &  \phn\dpm{250}{700}{250}  &  \phn0.84  &  \phn93 (100)  &  68\%     \\[0.5ex]
VW Hyi &  $\phn8.2\pm0.3$  &     $-0.05\pm0.06$  &  $\phn3.7\pm0.4$  &  $0.0+0.05$       &  \phn\dpm{580}{240}{250}  &  \phn0.81  &     181 (171)  &  29\%     \\[0.5ex]
WX Hyi &  $26.8\pm3.2$  &        $-0.72\pm0.05$  &  $\phn3.4\pm0.7$  &  $2.3\pm0.4\phn$  &  \phn\dpm{720}{310}{330}  &   7.1      &      82 (95)   &  82\%     \\[0.5ex]
SU UMa &  $19.3\pm1.8$  &  $\phm{-}0.07\pm0.06$  &      $140\pm20$   &  $1.2\pm0.2\phn$  &  \phn\dpm{430}{270}{430}  &  66.0\phn  &     256 (215)  &  \phn3\%  \\[0.5ex]
TY PsA &  $10.8\pm0.6$  &  $\phm{-}0.37\pm0.07$  &       $25\pm5$    &  $1.5\pm0.3\phn$  &     \dpm{1100}{500}{500}  &   4.8      &     143 (142)  &  45\%     \\[0.5ex]
AB Dra &  $21.6\pm1.8$  &        $-0.18\pm0.05$  &  $\phn3.7\pm0.6$  &  $6.1\pm0.4\phn$  &  \phn\dpm{660}{400}{660}  &   2.5      &     217 (196)  &  15\%     \\[0.5ex]
WW Cet &  $14.9\pm1.1$  &  $\phm{-}0.24\pm0.09$  &       $35\pm5$    &  $2.1\pm0.3\phn$  &  \phn\dpm{800}{460}{470}  &  10.7\phn  &     215 (192)  &  12\%     \\[0.5ex]
U Gem  &  $\phn55\pm10$ &        $-0.23\pm0.05$  &  $\phn2.6\pm0.4$  &  $0.5\pm0.2\phn$  &  \phn\dpm{730}{220}{210}  &   4.9      &     246 (209)  &  \phn4\%  \\[0.5ex]
EI UMa &  $50\pm5$      &  $\phm{-}0.05\pm0.05$  &     $13.0\pm0.6$  &  $0.7\pm0.4\phn$  &           --              &  17.4\phn  &     237 (241)  &  56\%     \\[0.5ex]
       &  $54\pm5$      &        $-0.03\pm0.05$  &     $12.6\pm0.7$  &  $1.3\pm0.2\phn$  &           --              &  19.8\phn  &     339 (241)  &  \phn0\%      
\enddata
\tablecomments{
The table shows the fit results for our cooling flow model described in Section~\ref{cflowmod}.
$\dot{M}_{bl}$ and $L_{bl}$ were derived using the distances in Table~\ref{nmsyspar} and with the
assumption that half of the belt-like boundary layer is obscured by the white dwarf.
The uncertainties represent the 90\% confidence level.
Note that the values for $\chi^2$ and the null hypothesis probability shown in the last two columns represent the fit quality
for the EPIC PN spectra only, whereas the fit parameters were obtained from combined fits to all EPIC and RGS spectra
(see Section~\ref{nmspec}).
Additional fit parameters for OY~Car and EI~UMa are shown in Table~\ref{nmfitresults2}.
}
\end{deluxetable}

\begin{deluxetable}{lccccccccc}
\tablecaption{Additional Fit Parameters\label{nmfitresults2}}
\tablewidth{0pc}
\rotate
\tabletypesize{\small}
\tablehead{
Object &  \multicolumn{2}{c}{Partial absorber}  &          Doppler          &  \multicolumn{2}{c}{Blackbody}  &                \multicolumn{4}{c}{Equivalent line widths (eV)}                          \\
       &          $N_H$        &   Covering     &      width $\sigma_v$     &  $kT_{bb}$ &        Area        &  O {\sc vii} {\it r}  &  O {\sc vii} {\it i}  &  O {\sc vii} {\it f}  &  Fe K-$\alpha$  \\
       &  ($10^{21}$cm$^{-2}$) &   fraction     &      (km$\,$s$^{-1}$)     &    (eV)    &      (km$^2$)      &        21.6~\AA       &        21.8~\AA       &         22.1~\AA      &     6.4~keV     }
\startdata
OY Car &  $9.9\pm1.2$         &  $0.49\pm0.03$  &             --            &     --     &        --          &           --          &           --          &           --          &       --        \\[1ex]
       &  $9.7\pm1.0$         &  $0.58\pm0.02$  &             --            &     --     &        --          &           --          &           --          &           --          &       --        \\[1ex]
EI UMa &  $36\pm5\phn$        &  $0.43\pm0.02$  &     \dpm{1140}{730}{300}  &  $50\pm5$  &      $13\pm9\phn$  &        $15\pm5$       &        $12\pm5$       &          $<3$         &  $\phn97\pm27$  \\[1ex]
       &  $35\pm4\phn$        &  $0.49\pm0.03$  &  \phn\dpm{900}{340}{320}  &  $52\pm5$  &      $19\pm13$     &        $20\pm4$       &        $12\pm4$       &          $<2$         &  $118\pm21$   
\enddata
\tablecomments{Additional parameters for our model fits to the spectra of OY~Car (Section~\ref{oycar})
and EI~UMa (Section~\ref{eiuma}).
}
\end{deluxetable}

\begin{deluxetable}{lcccccccc}
\tablecaption{Elemental Abundances\label{nmabund}}
\tablewidth{0pc}
\rotate
\tabletypesize{\small}
\tablehead{
Object &       C       &       N       &        O        &       Ne      &       Mg      &       Si      &       S       &        Fe       }
\startdata
T Leo  &  $0.6\pm0.8$  &  $0.9\pm0.9$  &  $0.79\pm0.12$  &  $0.5\pm0.3$  &  $1.0\pm0.4$  &  $0.9\pm0.2$  &  $0.5\pm0.3$  &  $0.72\pm0.07$  \\[0.5ex]
OY Car &  $2.2\pm1.1$  &  $0.0+0.9$    &  $0.53\pm0.12$  &  $0.8\pm0.3$  &  $0.8\pm0.4$  &  $1.4\pm0.2$  &  $1.3\pm0.3$  &  $1.57\pm0.08$  \\[0.5ex]
       &  $1.1\pm1.7$  &  $0.0+0.7$    &  $0.30\pm0.25$  &  $1.1\pm0.9$  &  $1.1\pm0.9$  &  $1.1\pm0.6$  &  $1.2\pm0.7$  &  $1.48\pm0.17$  \\[0.5ex]
VW Hyi &  $1.0\pm0.6$  &  $1.5\pm1.0$  &  $0.82\pm0.10$  &  $0.3\pm0.3$  &  $1.1\pm0.3$  &  $1.2\pm0.2$  &  $1.1\pm0.3$  &  $1.01\pm0.06$  \\[0.5ex]
WX Hyi &  $1.3\pm0.7$  &  $0.9\pm1.0$  &  $0.48\pm0.12$  &  $1.2\pm0.5$  &  $1.1\pm0.6$  &  $1.9\pm0.5$  &  $2.3\pm0.7$  &  $0.85\pm0.09$  \\[0.5ex]
SU UMa &  $1.2\pm0.6$  &  $0.4\pm0.6$  &  $0.41\pm0.08$  &  $0.3\pm0.2$  &  $0.7\pm0.3$  &  $0.8\pm0.2$  &  $0.3\pm0.3$  &  $0.58\pm0.06$  \\[0.5ex]
TY PsA &  $0.1\pm1.1$  &  $0.3\pm1.7$  &  $0.62\pm0.18$  &  $0.4\pm0.5$  &  $1.1\pm0.5$  &  $0.7\pm0.3$  &  $0.9\pm0.5$  &  $0.69\pm0.08$  \\[0.5ex]
AB Dra &  $1.1\pm0.7$  &  $0.0+0.3$    &  $0.28\pm0.08$  &  $0.2\pm0.2$  &  $0.7\pm0.3$  &  $0.6\pm0.2$  &  $0.5\pm0.3$  &  $0.45\pm0.05$  \\[0.5ex]
WW Cet &  $0.4\pm0.5$  &  $0.4\pm0.9$  &  $0.37\pm0.09$  &  $0.4\pm0.3$  &  $0.5\pm0.3$  &  $0.6\pm0.2$  &  $0.2\pm0.3$  &  $0.43\pm0.05$  \\[0.5ex]
U Gem  &  $0.0+0.4$    &  $5.1\pm1.3$  &  $0.72\pm0.11$  &  $0.1\pm0.4$  &  $1.5\pm0.5$  &  $1.7\pm0.3$  &  $0.8\pm0.4$  &  $1.40\pm0.10$  \\[0.5ex]
EI UMa &  $2.2\pm1.6$  &  $0.0+1.8$    &  $1.30\pm0.25$  &  $0.5\pm0.4$  &  $1.0\pm0.6$  &  $1.6\pm0.5$  &  $0.0+0.7$    &  $0.36\pm0.08$  \\[0.5ex]
       &  $1.0\pm1.3$  &  $0.6\pm1.6$  &  $1.00\pm0.19$  &  $0.1\pm0.3$  &  $1.3\pm0.5$  &  $1.2\pm0.4$  &  $0.0+0.3$    &  $0.45\pm0.07$  
\enddata
\tablecomments{
The table shows the elemental abundances obtained from our cooling flow model (Section~\ref{cflowmod}).
Abundances are given relative to the solar values in \citet{1989GeCoA..53..197A}.
The uncertainties represent the 90\% confidence level.
}
\end{deluxetable}

The results of our fits with the cooling flow model are shown in Tables~\ref{nmfitresults}--\ref{nmabund}.
The initial temperature $T_{max}$ of the flow covers a wide range from 8~keV for VW~Hyi to 55~keV for U~Gem.
We fixed the minimum temperature $T_{min}$ at the lower bound of our model (8~eV) because preliminary fits showed
that the parameter converges in all cases to this value.
The upper limits on $T_{min}$ derived from those fits are $\sim$200~eV for all objects
(at 90\% confidence level).
This is consistent with the accreting gas not becoming optically thick before it has cooled to the low
white dwarf temperature.
In most cases, the parameter $\alpha$ is near~0, demonstrating that the emission measure distributions are close
to that of the simple cooling flow model (Equation~\ref{emdist}).
Only for WX~Hyi, which was observed during outburst, does $\alpha$ indicate a significant deviation from
a simple cooling flow.
The apparently good quality of the fit is likely a result of the low signal-to-noise ratio of the WX~Hyi data.
To calculate the accretion rates $\dot{M}_{bl}$ and the unabsorbed luminosities $L_{bl}$, we used the distances
shown in Table~\ref{nmsyspar} and assumed that half of the belt-like boundary layer is obscured by the white dwarf.
We further assumed isotropic emission and neglected reflection of X-rays off the white dwarf.
Since typically $\sim$70\% of the flux is emitted in the 0.2--12~keV energy range and directly detected
by \xmm, the estimate of $L_{bl}$ is fairly model independent.

For most of the dwarf novae, we find significantly non-zero boundary layer rotation velocities $V_{bl}\sin{i}$.
However, all velocities are considerably smaller than the Keplerian velocity near the white dwarf's surface
($\sim$3000~km~s$^{-1}$).
The line broadening caused by the orbital motion of the white dwarf itself is of order 100~km~s$^{-1}$
and can be neglected.
Note that the resolution of the EPIC spectra is insufficient to resolve any Doppler broadening.
The rotation velocities are therefore solely constrained by the emission lines in the RGS spectra, 
in particular the strong O~{\sc viii} K-$\alpha$ line.
We found that fits to only the O~{\sc viii} K-$\alpha$ line yield similar results as the global fits,
albeit with slightly larger uncertainties.

The low values for $\chi^2$ demonstrate that the X-ray spectra are well described by our cooling flow model.
As mentioned earlier, the $\chi^2$ values in Table~\ref{nmfitresults} represent the fit quality for the
EPIC PN spectra only, while the fit parameters were obtained from combined fits to all EPIC and RGS spectra.
Using the $\chi^2$ of the EPIC PN spectra as a quality measure was necessary to avoid the bias
from the cross-calibration discrepancies between the instruments.
The elemental abundances relative to their solar values are shown in Table~\ref{nmabund}.
We obtained very good constraints for the abundances of O and Fe.
Not shown are the abundances for Na, Al, Ar, Ca, and Ni, which are included in the \mekal\ model,
but for which the fit did not provide useful constraints.


\subsection{OY~Car}
\label{oycar}

The spectrum of OY~Car is qualitatively similar to those of the other dwarf novae in our sample.
The only apparent difference is the lower flux at low energies, which may indicate intrinsic
absorption.
OY~Car, being the only eclipsing system in our sample, has a high orbital inclination ($83^\circ$),
and it is likely that some of the boundary layer emission is absorbed by the accretion disk.
We therefore included in our model absorption by a partially covering, neutral absorber
(\pcfabs\ model in \xspec).
This additional model component significantly improved the quality of the fit.
The best-fitting parameters for the partial absorber component are shown in Table~\ref{nmfitresults2}.

For the first of the two OY~Car observations, the somewhat large $\chi^2_{red}=1.45$ for the PN spectrum
seems to indicate a disagreement with our model.
However, we find that the combined MOS spectra are in very good agreement with our model
with $\chi^2_{red}=455/425=1.07$ (null hypothesis probability 15\%).
A comparison between the three EPIC spectra showed that below 0.5~keV the measured flux in the PN spectrum
is $\sim$10\% lower than in the MOS spectrum.
We therefore suspect that the large $\chi^2$ in Table~\ref{nmfitresults} is a result of calibration
uncertainties of the PN detector.
OY~Car is the only object in our sample that was observed early on during the \xmm\ mission
when instrument calibration was still at an early stage.
We also found discrepancies between the RGS and EPIC data at low energies.
Below 0.5~keV, the RGS measured more than double the flux seen with the EPIC detectors.
This discrepancy is the reason why our model fits for OY~Car appear to overpredict the EPIC flux
at low energies (see Fig.~\ref{nmepic}).

We conclude that the spectrum of OY~Car is in good agreement with our cooling flow model
provided that partial absorption by the accretion disk is considered.
In our model fits we used a simple partial absorber with a single column density and covering fraction.
In reality, absorption by the disk is likely more complex with a continuously varying column density.
However, the quality of our data is insufficient to fit models with multiple partial absorbers.


\subsection{EI~UMa}
\label{eiuma}

The spectrum of EI~UMa is qualitatively very different from those of the other objects in our sample.
The most noticeable differences are the harder spectral slope, the weakness of Fe L-shell emission at 0.7--1.3~keV,
the fluorescent Fe K-$\alpha$ line at 6.4~keV, and the strong O~{\sc vii} triplet at 22~\AA.
These spectral features are not typical for dwarf novae, but they suggest that EI~UMa is a magnetic system
of the intermediate polar type.
We will discuss this hypothesis further in Section~\ref{eiumadisc}.
For the purpose of spectral fitting, we will assume in this section that EI~UMa is an
intermediate polar.

The accretion disk in intermediate polars is disrupted at some radius by the magnetic field of the
white dwarf, and the disk material is funneled along the field lines onto the magnetic poles.
At some height above the white dwarf, the supersonic accretion flow forms a standoff shock that heats the gas
to temperatures in excess of $10^8$~K.
Below the shock, the hot plasma is compressed as it piles up on the white dwarf, and it begins to cool
via the emission of X-rays.
A recent review of the standard model for accretion onto magnetic white dwarfs is given in
\citet{2000SSRv...93..611W}.

We modeled the X-ray emission from the accretion column using the same cooling flow model as for the dwarf novae
(see Section~\ref{cflowmod}).
In intermediate polars, reflection of X-rays by the white dwarf slightly increases the observed flux
above $\sim$5~keV.
We therefore included a simple multiplicative reflection model (\hrefl\ in \xspec),
where we assumed an angle of 30$^\circ$ between the line of sight and the surface normal.
Note that the contribution from reflection is fairly small ($\sim$20\% at 10~keV),
and the use of a simple reflection model is sufficient.
The assumed reflection angle has only a small impact on our fit results,
and changing the angle to 60$^\circ$ only increases the two parameters of the partial absorber
component by $\sim$5\%.
Partial absorption of X-rays by the cooler pre-shock flow is observed in many intermediate polars.
Adding a partially covering, neutral absorber (\pcfabs\ in \xspec) to our model
significantly improved the quality of the fit.
The partial absorber was applied only to the cooling flow and none of the other components.

The cooling flow model does not account for the strong Fe K-$\alpha$ line at 6.4~keV and the O~{\sc vii}
line triplet at 22~\AA.
We therefore included these lines individually in our model.
The Fe K-$\alpha$ line is likely caused by fluorescence in the cooler pre-shock flow or the
photosphere of the white dwarf.
The observed flux in the energy range 0.2--0.3~keV is higher than predicted by the cooling flow model.
This flux excess is likely the high-energy tail of the thermal emission from the white dwarf's photosphere
below the accretion column.
To account for this excess, we added a single-temperature blackbody component to our model.
We further included Doppler broadening with a Gaussian line profile.

The fit results for our model are shown in Tables~\ref{nmfitresults} and~\ref{nmfitresults2}.
The parameter $\alpha$ is near zero, indicating that the post-shock flow is close to a
simple, isobaric cooling flow.
The high plasma temperature $kT_{max}\approx50$~keV is probably responsible for the harder spectral slope
compared to the dwarf novae.
The strong continuum emission from the high-temperature plasma considerably
reduces the equivalent widths of lines predominantly emitted by the cooler plasma in the accretion column.
This may explain the apparent weakness of the Fe L-shell lines.
We find that about half of the X-ray emitting region is obscured by a neutral absorber
with a large column density of $\sim$$4\times10^{22}$~cm$^{-2}$.
To calculate the total accretion rate $\dot{M}_{bl}$ and luminosity $L_{bl}$, we assumed that one of the two
accreting poles is obscured by the white dwarf.
The given $L_{bl}$ is the unabsorbed luminosity of the cooling flow component only.

The temperature of the blackbody component ($\sim$50~eV) is at the high end of the temperature range observed
in intermediate polars \citep[e.g.][]{1995A&A...297L..37H,2004A&A...415.1009D}.
The fractional emitting area of the blackbody $f_{eff}\approx3\times10^{-8}$ is very small,
considerably smaller than the typical size of the accretion region $f_{zone}\approx10^{-4}-10^{-3}$.
However, it is possible that $f_{eff}\ll f_{zone}$ if accretion is very inhomogeneous
and occurs via dense filaments or blobs \citep{1988MNRAS.235..433H}.
Blackbody emission with a similarly small effective area $f_{eff}$ has been observed in the X-ray spectra
of several other intermediate polars \citep{2002A&A...387..201H,2004A&A...415.1009D}.
Note that the bolometric luminosity of the blackbody component is only $\sim$0.5\% of the
cooling flow luminosity $L_{bl}$.
This suggests that considerably more blackbody radiation is emitted from a larger area at lower temperatures
and that our X-ray spectra only show emission from the hottest parts of the accretion region.

Our model is in excellent agreement with the spectrum from the first EI~UMa observation.
For the second observation, however, the model does not provide a good fit.
This is surprising as there are no apparent differences between the two observations other than a
slight difference in the X-ray and optical brightness.
We suspect that, during the second observation, absorption by the pre-shock accretion flow was more complex
than the simple partially covering, neutral absorber in our model.
Some improvement of the fit could be achieved by adding a second partial absorber to the model.
This may indicate a continuous distribution of column densities.
It is also possible that the absorbing material is partially ionized and the neutral absorber model is insufficient.
The differences between the two spectra may simply be a result of phase-dependent absorption.
However, our knowledge of the white dwarf spin period and the binary orbital period is insufficient
to compare the spin and orbital phases for the two observations.


\section{DISCUSSION}

Our sample of nine dwarf novae allows us to investigate some general properties of dwarf novae at
low accretion rates ($10^{-12}-10^{-10}$~$M_\odot$~yr$^{-1}$).
All of the dwarf novae, except WX~Hyi, were observed during their quiescent state.
Even though WX~Hyi was in outburst, its luminosity was lower than for some of the
quiescent dwarf novae.
EI~UMa, which is most likely an intermediate polar, is discussed separately in Section~\ref{eiumadisc}.


\subsection{X-ray/UV Luminosities and the Missing Boundary Layers}
\label{nmlumin}

\begin{figure}
\epsscale{0.5}
\plotone{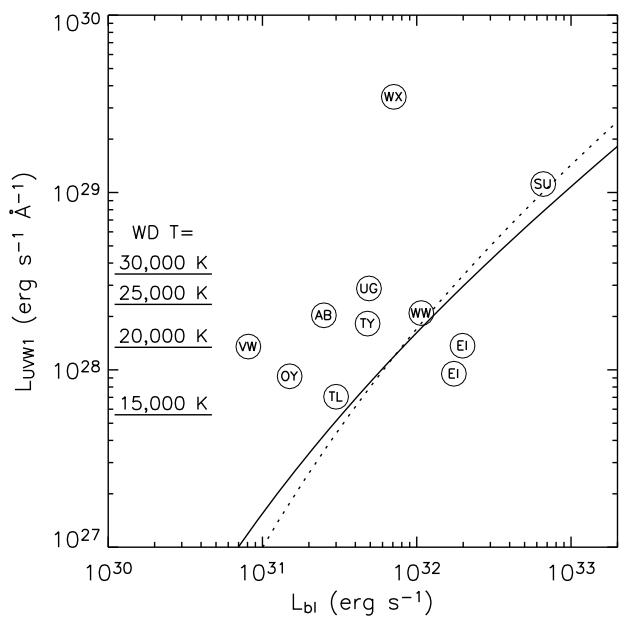}
\caption{\label{xuvlum}
Boundary layer luminosity derived from the X-ray spectrum vs.\ UV spectral luminosity at 290~nm
measured with the Optical Monitor (see Section~\ref{nmlumin}).
The solid and dotted lines show the UV luminosity predicted by a simple accretion disk model
for two inner disk radii $R_{in}=5,000$~km (solid line)
and $R_{in}=10,000$~km (dotted line).
The short horizontal lines show the UV luminosity of a white dwarf with radius
$R_{wd}=8,000$~km at various temperatures.
}
\end{figure}

Basic accretion theory predicts that roughly half of the gravitational
energy of the accreting matter is released by the disk as optical and UV radiation.
For white dwarfs rotating much slower than their breakup velocity, the other half should be
liberated as X-rays in the boundary layer, where the disk material is decelerated from
its Keplerian velocity to the rotation velocity of the white dwarf \citep[e.g.][]{1974MNRAS.168..603L}.
It is therefore expected that the disk luminosity $L_{disk}$ and the boundary layer luminosity $L_{bl}$
are roughly equal.
However, observations have shown that the X-ray luminosity is actually lower than predicted, which has become known
as the ``mystery of the missing boundary layers'' \citep[e.g.][]{1982ApJ...262L..53F,1994A&A...292..519V}.
In this section we investigate whether the boundary layers are underluminous
for the eight quiescent dwarf novae in our sample.

Fig.~\ref{xuvlum} shows the boundary layer luminosity $L_{bl}$ derived from the X-ray spectrum
vs.\ the spectral luminosity at 290~nm $L_{UVW1}$ obtained from the Optical Monitor data.
No OM data were available for AB~Dra, so we assumed a flux of 9~mJy, which we derived by scaling the spectrum
in \citet{1987A&AS...71..339V} to match the visual magnitude in the AAVSO light curve at the time
of the \xmm\ observation ($V=14.3$).
The solid and dotted lines in Fig.~\ref{xuvlum} show the $L_{UVW1}$ that is expected if $L_{disk}=L_{bl}$.
Here we derived $L_{UVW1}$ from $L_{disk}$ using a blackbody disk model
\citep[e.g.][]{1984PASJ...36..741M,1986ApJ...308..635M}.
The model requires only two parameters, the inner disk radius $R_{in}$ and the blackbody temperature at $R_{in}$.
Our predictions are shown for $R_{in}=5,000$~km (solid line) and $10,000$~km (dotted line),
which is appropriate for white dwarf masses $M_{wd}=1.1$~$M_\odot$ and 0.45~$M_\odot$, respectively
(assuming that $R_{in}=R_{wd}$).
Note that $M_{wd}$ and $R_{in}$ have only a small impact on the predicted $L_{UVW1}$.

For all dwarf novae in our sample, the observed $L_{UVW1}$ shown in Fig.~\ref{xuvlum} is at least equal to
or larger than the value expected if $L_{disk}=L_{bl}$.
Only the probable intermediate polar EI~UMa has a lower UV luminosity,
which is consistent with the truncation of the disk by the magnetic field (see Section~\ref{eiumadisc}).
The highest $L_{UVW1}$ was found for WX~Hyi, which was observed during outburst.
The UV luminosity in WX~Hyi was probably enhanced due to a high disk accretion rate.

\begin{figure}
\epsscale{0.5}
\plotone{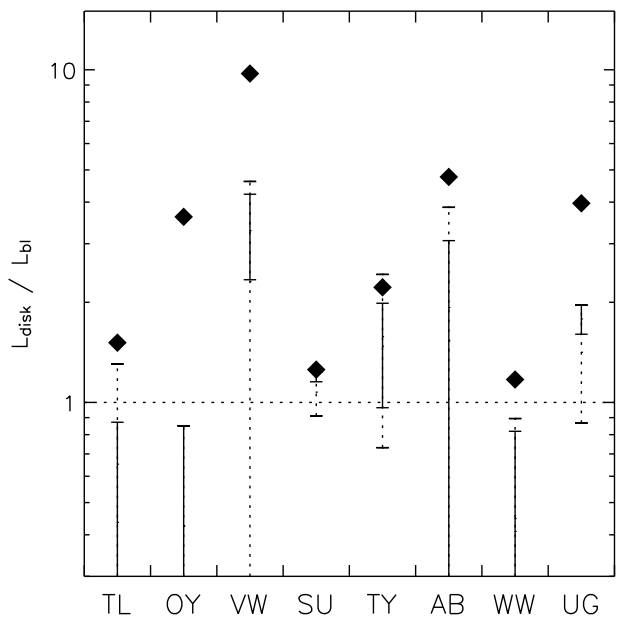}
\caption{\label{lbldisk}
Ratio of disk luminosity to boundary layer luminosity for the eight quiescent dwarf novae in our sample
(see Section~\ref{nmlumin}).
The diamond symbols show $L_{disk}/L_{bl}$ when all of the UV flux observed with \xmm\ is attributed to the disk.
Corrected luminosity ratios taking into account the UV emission from the white dwarf are shown as solid bars
for a reasonable range of surface temperatures.
The dotted vertical bars indicate how the possible range would be extended
if the disk were truncated at a radius up to $3R_{wd}$.
}
\end{figure}

The diamond symbols in Fig.~\ref{lbldisk} show the ratio $L_{disk}/L_{bl}$ for the eight quiescent dwarf novae.
The ratio was derived with the assumption that the observed UV flux is entirely due to the disk.
We estimated $L_{disk}$ from $L_{UVW1}$ using the blackbody disk model described above.
Here we assumed that $R_{in}=R_{wd}$.
Note that, even though accretion rates are fairly low,
the blackbody disk model describes the spectrum sufficiently well for our purposes.
While the outer parts of the disk may be optically thin, the inner regions,
where most of the accretion energy is released, are optically thick.
The calculations by \citet{1981AcA....31..127T} show that, for the accretion rates considered here,
the optical and UV continuum emission is well approximated by a blackbody disk model.
Emission or absorption lines do not significantly affect our results
since no major lines lie in the UVW1 bandpass.

The above derivation of $L_{disk}$ does not take into account the white dwarf emission,
which can be a major contributor to the UV flux.
The horizontal lines in Fig.~\ref{xuvlum} show the UV spectral luminosity of a white dwarf
at various temperatures.
To correct our estimates of $L_{disk}$, we assumed that the white dwarf radiates like a blackbody
and subtracted the white dwarf emission from $L_{UVW1}$ assuming a reasonable range of surface temperatures.
To calculate the visible emitting areas, we used the white dwarf radii and orbital inclinations shown
in Table~\ref{nmsyspar} and assumed that the surface below the equator is obscured by the disk.
For three of the dwarf novae, white dwarf temperatures have previously been measured, and we adopted the
following temperature ranges, which are appropriate for the number of days since the last outburst:
13,000--18,000~K for OY~Car \citep{1989A&A...213..167H,1994ApJ...426..294H},
18,000--20,000~K for VW~Hyi \citep{1996A&A...309L..47G}, and
31,000--34,000~K for U~Gem \citep{1994ApJ...424L..49L}.
For the dwarf novae without known white dwarf temperatures, we selected typical ranges based on the
distribution in \citet{1991AJ....102..295S,1999PASP..111..532S}:
12,000--20,000~K for orbital periods below the 2--3~hr gap (T~Leo, SU~UMa, and TY~PsA) and
20,000--40,000~K for periods above the gap (AB~Dra and WW~Cet).

The corrected $L_{disk}/L_{bl}$ ratios, obtained after subtracting the white dwarf emission,
are shown in Fig.~\ref{lbldisk} as solid vertical bars.
For six of the dwarf novae, the allowed range is consistent with $L_{disk}\approx L_{bl}$.
Only for VW~Hyi and U~Gem does $L_{disk}/L_{bl}$ appear to be notably above~1.
It has been suggested that the disk in quiescent dwarf novae is truncated at a radius
larger than $R_{wd}$ \citep[e.g.][]{1997MNRAS.288L..16K,1995A&A...302L..29L}.
Truncation of the disk would impact our estimates of $L_{disk}$ because it changes the parameter $R_{in}$
of the disk model and the fraction of the white dwarf's surface obscured by the disk.
The dotted vertical bars in Fig.~\ref{lbldisk} show how the allowed range of $L_{disk}/L_{bl}$
would be extended if the disk were truncated somewhere between $R_{wd}$ and $3R_{wd}$.
The extended ranges for VW~Hyi and U~Gem are now also consistent with $L_{disk}\approx L_{bl}$.
Note, however, that in a truncated disk less gravitational energy is available,
and one might expect that $L_{disk}/L_{bl}<1$.

We conclude that the X-ray and UV emission from the eight quiescent dwarf novae is consistent
with $L_{disk}\approx L_{bl}$ if the white dwarf emission and a possible truncation of the disk
are taken into account.
An excess of disk emission can be ruled out for four dwarf novae (T~Leo, OY~Car, SU~UMa, and WW~Cet),
though an excess of boundary layer luminosity is possible for three of them.
For the other four dwarf novae (VW~Hyi, TY~PsA, AB~Dra, and U~Gem), an underluminous boundary layer
may be present, but the $L_{disk}/L_{bl}$ ratio can be at most 2--4.
It is evident that in quiescent dwarf novae the problem of the underluminous boundary layers,
if it exists at all, is much less pronounced than in dwarf novae during outburst.


\subsection{Accretion Rates during Quiescence}

\begin{figure}
\epsscale{0.5}
\plotone{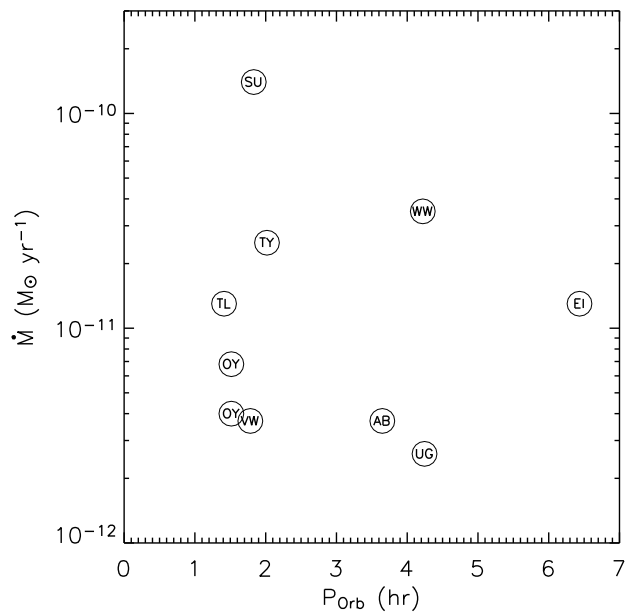}
\caption{\label{porbmdot}
Boundary layer accretion rates vs.\ orbital periods for the quiescent objects in our sample.
}
\end{figure}

Fig.~\ref{porbmdot} shows the boundary layer accretion rates vs.\ the orbital periods for
the quiescent objects in our sample.
The accretion rates cover a range of almost 2~orders of magnitude.
The scatter of data points is similar to that found by \citet{1987MNRAS.227...23W} for the absolute
visual magnitudes of the accretion disks in dwarf novae at minimum light.
A correlation with the orbital period or the outburst interval as found by \citet{1987MNRAS.227...23W}
is not apparent in our data.
Note that accretion rates derived from the X-ray spectrum are fairly reliable
and, unlike those estimated from the disk luminosity,
do not require a detailed understanding of the disk physics or knowlege of the orbital inclination.


\subsection{Boundary Layer Rotation Velocities}
\label{velocities}

\begin{figure}
\epsscale{0.5}
\plotone{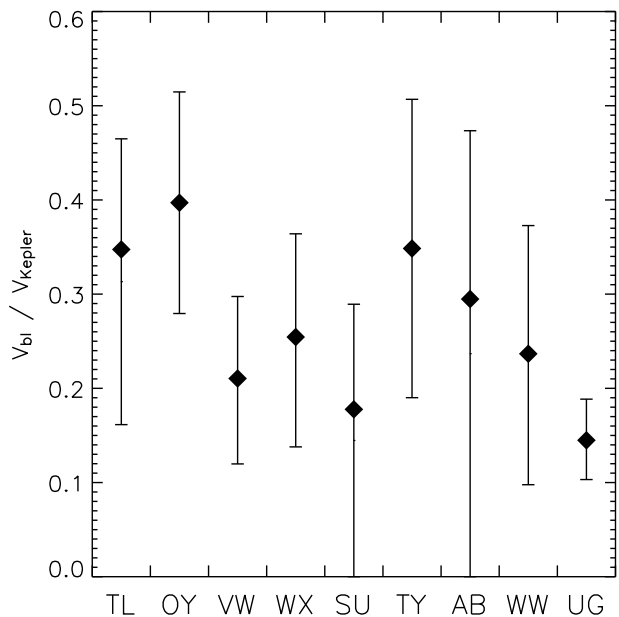}
\caption{\label{vrot}
Ratio of boundary layer rotation velocity $V_{bl}$ and Keplerian velocity $V_{Kepler}$ at the white dwarf radius
for the nine dwarf novae in our sample (see Section~\ref{velocities}).
Error bars have been derived from the 90\% confidence ranges of $V_{bl}\sin{i}$ in Table~\ref{nmfitresults}
and do not include the uncertainties of $i$ and $V_{Kepler}$.
}
\end{figure}

Fig.~\ref{vrot} shows the ratio of the boundary layer rotation velocities $V_{bl}$
(derived from $V_{bl}\sin{i}$ in Table~\ref{nmfitresults})
and the Keplerian velocities at the white dwarf radius $V_{Kepler}=\sqrt{GM_{wd}/R_{wd}}$.
Note that the error bars in Fig.~\ref{vrot} only include the uncertainty of the measured $V_{bl}\sin{i}$
and not that of $i$ or $V_{Kepler}$.
Since approximately $V_{Kepler}\propto M_{wd}$$^{0.9}$, the uncertainty of $V_{bl}/V_{Kepler}$ is somewhat larger
for systems with poorly known white dwarf masses.

It is evident from Fig.~\ref{vrot} that, for all dwarf novae in our sample, the boundary layer rotates
considerably slower than the inner accretion disk.
Note, however, that the measured $V_{bl}$ depends mostly on the width of the O~{\sc viii} K-$\alpha$ line
and is therefore an estimate of the rotation velocity of the cooler gas in the lower parts of the boundary layer.
In a cooling flow with $kT_{max}=20$~keV, half of the O~{\sc viii} line flux
is emitted at plasma temperatures below 2~keV.
The white dwarf rotation velocity $V_{wd}$ is likely lower than the rotation velocity of the boundary layer,
and $V_{bl}$ can be considered an upper limit on $V_{wd}$.
Consequently, all dwarf novae in our sample contain a white dwarf that is rotating considerably below the
breakup velocity $V_{Kepler}$.

White dwarf rotation velocities have been measured for three objects in our sample.
In VW~Hyi, the white dwarf is rotating with a projected velocity $V_{wd}\sin{i}=400-500$~km~s$^{-1}$
\citep{2001ApJ...561L.127S}, which is close to our measurement of the boundary layer velocity
$V_{bl}\sin{i}\approx580$~km~s$^{-1}$.
In contrast, the upper limits on $V_{wd}\sin{i}$ of $\le$100~km~s$^{-1}$ for U~Gem \citep{1994ApJ...430L..53S}
and $<$200~km~s$^{-1}$ for OY~Car \citep{1994AIPC..308..197C} are significantly smaller than
the boundary layer velocities of $\sim$730~km~s$^{-1}$ and  $\sim$1350~km~s$^{-1}$, respectively.
It is apparent that a high $V_{bl}$ does not necessarily imply a high $V_{wd}$,
and $V_{wd}$ for the other dwarf novae could be anywhere in the range 0 to $V_{bl}$.


\subsection{Cooling Flow Temperatures}
\label{temperatures}

\begin{figure}
\epsscale{0.5}
\plotone{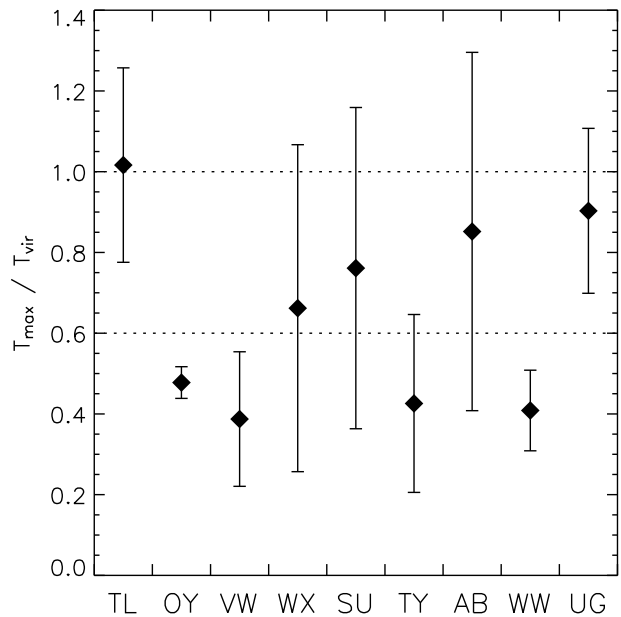}
\caption{\label{tmax}
Ratio of initial cooling flow temperature $T_{max}$ and virial temperature $T_{vir}$
for the nine dwarf novae in our sample (see Section~\ref{temperatures}).
The two horizontal lines indicate ratios of 1 and $3/5$, respectively.
Error bars show the 90\% confidence range of $T_{max}$ combined with the uncertainty
of the white dwarf mass $M_{wd}$.
For systems without known masses (SU~UMa, TY~PsA, AB~Dra), we assumed $M_{wd}=0.7\pm0.2$~$M_\odot$
and for T~Leo $M_{wd}=0.4\pm0.05$~$M_\odot$.
}
\end{figure}

The initial cooling flow temperatures $T_{max}$ of the nine dwarf novae cover a wide range
from $\sim$8~keV for VW~Hyi to $\sim$55~keV for U~Gem with most temperatures concentrated
around 10--20~keV (Table~\ref{nmfitresults}).
The exceptionally high temperature in U~Gem is likely a result of the larger white dwarf mass.

In Fig.~\ref{tmax} we compare $T_{max}$ to the virial temperature $T_{vir}$ at the white dwarf radius.
$T_{vir}$ is the temperature that the disk material would have if all the kinetic energy
from its Keplerian motion were instantly converted into heat.
The virial temperature is given by $3/2\,kT_{vir}=1/2\,\mu m_p V_{Kepler}^2$, where $m_p$ is the proton mass
and $\mu$ the mean molecular weight ($\sim$0.6).
Since this equation gives the maximum energy per particle that can be dissipated,
the temperature in the boundary layer cannot exceed $T_{vir}$.
Indeed, $T_{max}\le T_{vir}$ appears to be satisfied for our sample of dwarf novae.

However, energy considerations for a cooling flow suggest that $T_{max}$ is somewhat smaller than $T_{vir}$.
As discussed in Section~\ref{cflowmod}, an isobaric cooling flow releases an energy of $5/2\,kT_{max}$ per particle,
$1\,kT_{max}$ of which comes from the pressure work required to compress the gas.
Because only an energy of $3/2\,kT_{vir}$ is available at the inner disk edge,
one might expect that $T_{max}=3/5\,T_{vir}$.
For the majority of objects in our sample, the data are consistent with $T_{max}/T_{vir}\le3/5$.

Several processes can lead to a reduced maximum temperature in an optically thin boundary layer.
If the X-ray emitting gas is rotating with a significant velocity,
not all of the kinetic energy has been dissipated when the plasma begins to radiate,
and $T_{max}$ will be reduced.
It is also possible that a significant amount of heat is conducted away from the hottest regions
of the boundary layer before the plasma can radiate efficiently in X-rays.
However, heat conduction does not dominate over cooling
via X-ray emission since otherwise $T_{max}$ would be much smaller than $T_{vir}$.
Furthermore, some of the accretion energy could be carried away in a wind,
which would reduce the energy available to heat the remaining gas.


\subsection{Structure of the Boundary Layer}
\label{blstruct}

The most striking result of our analysis is the excellent agreement of the X-ray spectra with
a simple, isobaric cooling flow.
This strongly suggests that the observed X-ray emission is produced by cooling plasma
settling onto the white dwarf.
The presence of a cooling flow has important implications for models of accretion in dwarf novae.
In particular, it seems unlikely that a major fraction of the X-rays is emitted
by an extended corona.
If such a corona is present, its density must be too low for the emission of significant amounts
of X-rays.

The high initial temperature of the cooling flow, which is close to the virial temperature
(Section~\ref{temperatures} and Fig.~\ref{tmax}),
indicates that the accreting gas in the boundary layer does not cool substantially
until it has dissipated most of its rotational kinetic energy.
It is apparent that the density in the upper parts of the boundary layer is insufficient for the
emission of significant amounts of X-rays.

The results of our spectral analysis suggest the following picture of the boundary layer.
At the inner disk edge, the accreting gas is decelerated well below the Keplerian velocity,
and rotation can no longer support it against radial infall.
The gas is accelerated towards the white dwarf while it continues to lose angular momentum
until most of its rotational kinetic energy has been dissipated.
This dissipation heats the accretion flow to temperatures comparable to the virial temperature
($\ge$$10^8$~K).
The inward acceleration as well as a vertical expansion of the boundary layer, caused by the
large temperature increase, leads to a considerable drop in density.
Because of the low density, the gas can no longer radiate efficiently and becomes invisible to us.
Eventually, when the hot accretion flow begins to pile up on the white dwarf, the density increases rapidly,
and the plasma begins to cool strongly via the emission of X-rays while closely resembling an isobaric
cooling flow.
This picture of the boundary layer is in remarkable qualitative agreement with the model
calculations by \citet{1993Natur.362..820N} for dwarf novae at low accretion rates.

An alternative picture of the inner accretion disk is discussed in \citet{1994A&A...288..175M}.
The authors suggested that the inner disk in quiescent dwarf novae becomes unstable and is evaporated
by a coronal siphon flow.
The gas in the forming corona is partially accreted onto the white dwarf and partially lost in a wind.
If the density in the corona is low, the X-ray spectrum would be dominated by emission from the
cooling plasma piling up on the white dwarf.
This scenario is also consistent with the results of our spectral analysis.

The two pictures of the boundary layer discussed above are also supported by our timing analysis
of VW~Hyi data presented in \citet{2003MNRAS.346.1231P}.
We discovered that the variations of the X-ray and the UV flux seen on a time scale of $\sim$1500~s
are correlated and that the X-ray fluctuation are delayed with respect to the UV by $\sim$100~s.
We attributed this correlation to accretion rate fluctuations that propagate from
the inner disk to the X-ray emitting part of the boundary layer.
However, no such correlations were found for the other dwarf novae in our sample.


\subsection{Elemental Abundances}
\label{abunddisc}

The elemental abundances in the boundary layer are shown in Table~\ref{nmabund}.
We obtained good constraints for O and Fe as these elements
exhibit strong Ly-$\alpha$ and K-$\alpha$ emission lines (Figs~\ref{nmepic}--\ref{nmrgs2}).
Somewhat less accurate are our abundance estimates for Ne, Mg, Si, and S,
while only poor constraints were derived for C and N.
Our results are mostly consistent with near solar abundances.
One noticeable exception is the high nitrogen abundance in U~Gem (see below).
The oxygen abundances are consistently subsolar for the dwarf novae with an average
value of $\sim$0.5~solar.
For the low to moderate accretion rates during our observations, it is unlikely
that significant mixing between the boundary layer and the white dwarf occurred.
Our spectral analysis therefore measures the elemental abundances in the accreting matter,
which are equal to the abundances on the companion star.

For two of the dwarf novae in our sample, the abundances in the white dwarf's photosphere have been
determined from UV spectroscopic observations.
\citet{1999ApJ...511..916L} analyzed white dwarf spectra of U~Gem and found an overabundance of nitrogen
at 4 times solar and an underabundance of carbon at 0.1 times solar.
The authors explain these distinctly nonsolar abundances via accretion of material that has previously
been processed by CNO burning, either in the core of a massive star or in a weak nova explosion on the
white dwarf.
From our spectral analysis of the U~Gem data, we find similar nonsolar abundances in the accretion flow
($\sim$5 for nitrogen and $<$0.4 for carbon).
This confirms that the origin of the unusual composition on the white dwarf is indeed accretion
of material from a secondary with nonsolar abundances.
Interestingly, none of the other objects in our sample show a similar overabundance of nitrogen.
The unusual abundances on the secondary in U~Gem may be the result of a more massive primary
in the pre-cataclysmic binary.
During common envelope evolution, the composition of the secondary can be significantly altered
by the mass transfer from the red giant primary.
If the primary is massive enough to allow CNO burning, this contamination of the secondary
could lead to unusual nitrogen and carbon abundances as those observed in U~Gem.
A primary more massive than in most pre-cataclysmic binaries is indicated by the comparatively
high mass of the white dwarf in U~Gem.

White dwarf photospheric abundances indicative of CNO burning were also found in VW~Hyi
\citep{2001ApJ...561L.127S,1997ApJ...480L..17S}.
The authors interpreted this as evidence that a thermonuclear runaway has occurred on the white dwarf.
Our spectral analysis of the \xmm\ data did not reveal such distinctly non-solar abundances in the boundary layer.
This clearly demonstrates that the unusual abundances in VW~Hyi are indeed a result of nuclear burning on the
white dwarf and not accretion from the secondary.

Both a thermonuclear runaway on the white dwarf and accretion from the secondary
could be responsible for the unusual abundances observed in some dwarf novae.
An X-ray spectral analysis with a cooling flow model can be an important tool to measure the elemental abundances
on the secondary and therefore distinguish between the two possibilities.


\subsection{EI~UMa}
\label{eiumadisc}

The spectrum of EI~UMa differs significantly from those of the other objects in our sample
(see Section~\ref{eiuma}).
The most distinctive features are the harder spectral slope, the weakness of Fe L-shell emission at 0.7--1.3~keV,
the fluorescent Fe K-$\alpha$ line at 6.4~keV, and the strong O~{\sc vii} triplet at 22~\AA.
These spectral features are typically not seen in dwarf novae but appear to be prominent in
intermediate polars.
In a study of \chandra\ HETG spectra of four intermediate polars and three dwarf novae \citep{2003ApJ...586L..77M},
three of the intermediate polars exhibited the same distinctive spectral features as EI~UMa
(harder spectral slope, weak Fe L-shell lines, strong O~{\sc vii} lines),
whereas the three dwarf novae had cooling flow spectra similar to the dwarf novae in our sample
with strong Fe L-shell emission and only weak O~{\sc vii} lines.
A prominent feature in the spectrum of EI~UMa is the fluorescent Fe K-$\alpha$ line at 6.4~keV,
which appears to be strong in magnetic cataclysmic variables \citep{1999ApJS..120..277E} but is weak or absent
for the nine dwarf novae in our sample.
Because of the distinctive spectral features, we suspect that EI~UMa is a magnetic cataclysmic variable
of the intermediate polar type.

In the catalog by \citet{2003A&A...404..301R}, EI~UMa is classified as a U~Gem type dwarf nova.
However, we were not able to find published results that support this classification, and we suspect that it
is based solely on the non-detection of X-ray oscillations caused by the white dwarf rotation
\citep{1985MNRAS.215P..81C}.
To our knowledge, no dwarf nova outbursts, which would justify this classification, have been observed.
\citet{1986AJ.....91..940T} suggested that EI~UMa may be a DQ~Her star (intermediate polar) based on
the X-ray characteristics and strong He~{\sc ii} $\lambda$~4686 \citep{1982PASP...94..560G}.
We searched the \xmm\ data for periodic oscillations that would identify EI~UMa as an intermediate polar,
but we could not find a clearly periodic signal in either the X-ray or the UV light curves.
However, at periods longer than $\sim$100~s, the periodograms displayed strong spectral power due to the flickering
common in cataclysmic variables, which could mask small periodic oscillations.
Since the orbital inclination of EI~UMa is low \citep[$\sim$$23^\circ$;][]{1986AJ.....91..940T},
the accretion region is probably not occulted by the white dwarf, and the oscillation amplitude
is likely to be small.
If the white dwarf is rotating slowly (period $>$100~s), periodic oscillations due to the white dwarf rotation
may therefore be difficult to detect.
In intermediate polars, there appears to be a rough correlation between the spin and orbital periods.
For the 6.4-hr orbital period of EI~UMa, the sample of intermediate polars in \citet{1991MNRAS.248..370W}
suggests a white dwarf spin period $>$2000~s.
The duration of the two \xmm\ observations ($<$10000~s) is insufficient to detect small 
oscillations with such long periods.
We therefore conclude that the non-detection of oscillations does not rule out the possibility that EI~UMa
is a magnetic system.

The strongest evidence for EI~UMa being an intermediate polar is probably the low observed UV luminosity.
As shown in Fig.~\ref{xuvlum}, EI~UMa is the only object in our sample with a $L_{UVW1}$
lower than predicted from the assumption of equal disk and boundary layer luminosities.
It is difficult to explain such a low $L_{UVW1}$ if EI~UMa were a dwarf nova.
However, if EI~UMa is an intermediate polar, the accretion disk will be disrupted by the white dwarf's magnetic field,
and the low UV luminosity is simply the result of a missing inner disk.
We estimated the inner disk radius $R_{in}$ with the relation $L_{disk}=GM_{wd}\dot{M}_{bl}/R_{in}$.
Using a simple disk blackbody model to derive $L_{disk}$ from $L_{UVW1}$,
we obtained an inner disk radius of 2.6~$R_{wd}$ for the first and 2.0~$R_{wd}$ for the second EI~UMa observation.
From a simple magnetic dipole model \citep[e.g.][Chapter 6.3]{1992apa..book.....F}, we estimate a surface magnetic field
strength of $1\times10^4$~G, which is at the lower end of field strengths in intermediate polars.

According to the standard model of the accretion column in magnetic white dwarfs, the shock temperature is
directly related to the white dwarf mass \citep[e.g.][]{2000SSRv...93..611W}.
With the assumption that the shock temperature is the same as the initial cooling flow temperature of $\sim$50~keV
(i.e.\ the post-shock gas does not cool until it is dense enough to radiate efficiently in X-rays),
we derive a white dwarf mass of 0.95~$M_\odot$.
Here it was also assumed that the shock height is small compared to the white dwarf radius.

The presence of a partially covering absorber in EI~UMa provides further support for our intermediate polar classification.
Of the nine dwarf novae in our sample, only the eclipsing binary OY~Car shows evidence of such an absorber.
If EI~UMa were a dwarf nova, its low orbital inclination of $\sim$$23^\circ$ would make the presence of
a partially covering absorber unlikely.
For intermediate polars, however, partial absorption by the cool gas in the pre-shock accretion column is widely observed
\citep[e.g.][]{1999ApJS..120..277E}.

The equivalent width of $\sim$110~eV of the fluorescent Fe K-$\alpha$ line at 6.4~keV is similar to those
found in intermediate polars, which are typically around 100--150~eV \citep{1999ApJS..120..277E}.
The likely origin of the fluorescent emission is the cool pre-shock accretion flow or the white dwarf's
photosphere near the accretion region.
Because the accretion flow is responsible for both the partial absorption and some of the fluorescence,
one might expect a connection between the column density and the equivalent width of the fluorescent Fe K-$\alpha$ line.
Using the method described in \citet{1999ApJS..120..277E}, we estimate that the pre-shock flow contributes
$\sim$10~eV to the equivalent line width.
However, this result is somewhat uncertain since we only know the properties of the partial absorber
along the line of sight.
The fluorescent Fe K-$\alpha$ emission in EI~UMa is probably dominated by the white dwarf.
The calculations by \citet{1991MNRAS.249..352G} for reflection off a circular slab of cold material
predict an equivalent line width of 110~eV for a viewing angle of $60^\circ$ with respect to the surface normal.

The strenghts of the resonance (r), intercombination (i), and forbidden (f) lines of the O~{\sc vii}
triplet at 22~\AA\ (Table~\ref{nmfitresults2}) can be used to determine the temperature and density in the region
where oxygen is predominantly in a He-like ionization state.
The line ratios widely used for He-like ions are $G=(f+i)/r$ and $R=f/i$
\citep[e.g.][and references therein]{2000A&AS..143..495P}.
For EI~UMa these ratios are $G=0.7$ and $R<0.2$.
The low value of $G$ indicates that collisional ionization and not photoionization is the dominant process
in the plasma.
It also shows that the temperature of the gas emitting the O~{\sc vii} lines is $>$200~eV
\citep[see calculations in][]{2000A&AS..143..495P}.
The low $R$-ratio (weak forbidden line) seems to suggest a high electron density $n_e\ge10^{12}$~cm$^{-3}$.
However, a low $R$-ratio can also be caused by a strong UV radiation field.
As shown in \citet{2002pcvr.conf..113M}, the radiation from a 30000-K blackbody is sufficient for the $R$ ratio
of O~{\sc vii} to be in the high density limit, even if the density is low.
Since the photosphere below the accretion column typically has temperatures this high,
the $R$-ratio cannot be used as a density diagnostic.

The origin of the strong O~{\sc vii} line emission is somewhat puzzling.
The overall spectrum is well described by our cooling flow model, yet the integrated O~{\sc vii} line flux is
about a factor of 10 larger than predicted by this model.
One might suspect that the emission measure at the bottom of the accretion column, where temperatures are low enough
for O~{\sc vii} to be the dominant ion, is larger than predicted by the cooling flow model.
However, if this were the case, the temperature obtained from the $G$-ratio ($>$200~eV) would imply a stronger
O~{\sc viii} K-$\alpha$ line than is actually observed.

We suspect that the O~{\sc vii} line emission originates from a transient plasma near the shock.
A transient plasma can be present if the electron temperature is changed so quickly that the population densities
of the different ionization states do not have enough time to reach their equilibrium values
\citep[for a discussion of transient plasmas see][]{1999xrsa.conf..189L}.
Such a fast temperature change occurs at the accretion shock in magnetic cataclysmic variables.
Before the shock, the accreting gas is cool, and oxygen is in a low ionization state.
At the shock, the electron temperature rises quickly to several tens of keV, but the atoms will remain in
lower ionization states for some time before they can be ionized to O~{\sc viii}.
The X-ray spectral signature of such a transient plasma would be a continuum consistent with a high electron temperature
and greatly enhanced O~{\sc vii} line emission.

In a transient plasma, O~{\sc vii} will be the dominant ionization state until a time $t$ after the shock given by
$n_e\,t\approx10^{11}$~cm$^{-3}$~s \citep{1999xrsa.conf..189L}.
For a typical electron density $n_e\approx10^{13}$~cm$^{-3}$ \citep{2002pcvr.conf..113M},
$t$ is $\sim$0.01~s.
Since the post-shock velocity is $\sim$1000~km~s$^{-1}$ ($1/4$ of free-fall velocity), the region in which
oxygen line emission is dominated by O~{\sc vii} extends to $\sim$10~km or $\sim$0.002~$R_{wd}$ below the shock.
For a small shock height, this can be a significant fraction of the accretion column.
We conclude that a transient plasma is a viable explanation for the strong O~{\sc vii} lines observed in EI~UMa.
With this interpretation it also becomes evident why the O~{\sc vii} line triplet is much weaker in dwarf novae
where no single standoff shock is present.
The many smaller shocks in the turbulent boundary layer of a dwarf nova heat the gas much more slowly
than the standoff shock in a magnetic system.
Note that in a transient plasma the He-like lines of elements other than oxygen should also be enhanced.
However, we expect these He-like lines to be very weak since we do not detect the Ly-$\alpha$ lines
of any elements other than oxygen in the RGS spectra.


\section{CONCLUSIONS}
\label{conclusions}

Our spectral analysis of \xmm\ data for eight quiescent dwarf novae revealed that the X-ray emission
originates from a hot, optically thin multi-temperature plasma.
The temperature distribution in the plasma is in close agreement with an isobaric cooling flow,
which points to a cooling plasma settling onto the white dwarf as the source of the X-rays.
We found that the initial temperature of the cooling flow is comparable to the virial temperature.
This is an indication that the accreting gas in the boundary layer does not cool substantially until
it has dissipated most of its kinetic energy and begins to settle onto the white dwarf.
We found that the X-ray emitting gas in the cooler parts of the boundary layer is
moving significantly slower than the inner accretion disk.
This implies that the white dwarfs of all dwarf novae in our sample are rotating considerably
slower than their breakup velocity.

Our findings are in qualitative agreement with the model calculations by \citet{1993Natur.362..820N}
for a disk-like boundary layer.
Conversely, the presence of an extended, X-ray emitting corona can be excluded.
With the new generation of X-ray telescopes, our observational knowledge of the boundary layer
has reached a level of detail that needs to be matched by theoretical model calculations.
Of special interest to the observer is the predicted distribution of emission measure vs.\ temperature
for the X-ray emitting gas.
In particular, it would be useful to know how theoretical predictions compare to the simple, isobaric cooling flow
that appears to be the dominant source of X-rays in dwarf novae.

We found no evidence of an underluminous boundary layer in any of the eight quiescent dwarf novae,
and our data are consistent with $L_{disk}\approx L_{bl}$ in all cases.
For four of the dwarf novae, $L_{disk}/L_{bl}>1$ can be clearly ruled out, whereas for the others
an underluminous boundary layer could be present with $L_{disk}/L_{bl}$ ratios up to 2--4.
Our findings disagree with those of \citet{1994A&A...292..519V}, who measured UV to X-ray flux ratios
of 10 and higher.
However, the authors did not consider the contribution from the white dwarf, which may dominate
over the disk emission in the UV.
Our results seem to suggest that there is no actual ``mystery of the missing boundary layers''
for the majority of quiescent dwarf novae.
The key to understanding whether there really is a missing boundary layer problem
is an accurate determination of the true disk luminosity.

We determined the elemental abundances in the boundary layer, which are likely equal to the abundances on
the companion star.
With a few exceptions, abundances are close to solar values.
We have demonstrated on two of the dwarf novae that X-ray spectra can be used to identify
the origin of unusual abundances on the white dwarf.
We found that in VW~Hyi the unusual abundances are the result of a thermonuclear runaway
on the white dwarf, whereas in U~Gem they were caused by accretion from the companion star.
For most of the dwarf novae, absorption of boundary layer emission by the accretion disk does not appear to
be important.
Only for the eclipsing dwarf nova OY~Car, which has a high orbital inclination of $83^\circ$,
do we find evidence of a partially covering absorber.

The X-ray spectrum of EI~UMa differs considerable from those of the other objects in our sample.
The most distinctive features are the weakness of Fe L-shell emission, the strong fluorescent Fe K-$\alpha$ line,
and the strong O~{\sc vii} line triplet.
We also found that the disk luminosity is significantly lower than the X-ray luminosity.
Our findings provide ample evidence that EI~UMa is probably an intermediate polar and not as previously though
a U~Gem type dwarf nova.
The high shock temperature of $\sim$50~keV indicates a white dwarf mass of $\sim$0.95~$M_\odot$.
The strong O~{\sc vii} line triplet is possibly a result of a transient plasma below the shock.
We have demonstrated that X-ray spectroscopic observations can be used to distinguish intermediate polars
from dwarf novae.


\acknowledgments

This work is based on observations obtained with \xmm, an ESA science mission
with instruments and contributions directly funded by ESA member states and the
USA (NASA).
DP and FAC acknowledge support from NASA grant NAG5-12390.
We acknowledge with thanks the variable star observations from the AAVSO International Database
contributed by observers worldwide and used in this research.


\bibliographystyle{apj}
\bibliography{nmcvs}

\begin{thebibliography}{71}
\expandafter\ifx\csname natexlab\endcsname\relax\def\natexlab#1{#1}\fi

\bibitem[{{Anders} \& {Grevesse}(1989)}]{1989GeCoA..53..197A}
{Anders}, E., \& {Grevesse}, N. 1989, \gca, 53, 197

\bibitem[{{Arnaud}(1996)}]{1996adass...5...17A}
{Arnaud}, K.~A. 1996, in ASP Conf. Ser. 101, Astronomical Data Analysis
  Software and Systems V, ed. G.~Jacoby \& J.~Barnes (San Francisco: ASP), 17

\bibitem[{{Bruch} {et~al.}(1996){Bruch}, {Beele}, \&
  {Baptista}}]{1996A&A...306..151B}
{Bruch}, A., {Beele}, D., \& {Baptista}, R. 1996, \aap, 306, 151

\bibitem[{{Cash}(1979)}]{1979ApJ...228..939C}
{Cash}, W. 1979, \apj, 228, 939

\bibitem[{{Cheng} {et~al.}(1994){Cheng}, {Marsh}, {Horne}, \&
  {Hubeny}}]{1994AIPC..308..197C}
{Cheng}, F.~H., {Marsh}, T.~R., {Horne}, K., \& {Hubeny}, I. 1994, in AIP Conf.
  Proc. 308, The Evolution of X-ray Binaries, ed. S.~S. Holt \& C.~S. Day (New
  York: AIP), 197

\bibitem[{{Cook}(1985)}]{1985MNRAS.215P..81C}
{Cook}, M.~C. 1985, \mnras, 215, 81P

\bibitem[{{de Martino} {et~al.}(2004){de Martino}, {Matt}, {Belloni}, {Haberl},
  \& {Mukai}}]{2004A&A...415.1009D}
{de Martino}, D., {Matt}, G., {Belloni}, T., {Haberl}, F., \& {Mukai}, K. 2004,
  \aap, 415, 1009

\bibitem[{{den Herder} {et~al.}(2001)}]{2001A&A...365L...7D}
{den Herder}, J.~W., {et~al.} 2001, \aap, 365, L7

\bibitem[{{Ezuka} \& {Ishida}(1999)}]{1999ApJS..120..277E}
{Ezuka}, H., \& {Ishida}, M. 1999, \apjs, 120, 277

\bibitem[{{Ferland} {et~al.}(1982){Ferland}, {Pepper}, {Langer}, {MacDonald},
  {Truran}, \& {Shaviv}}]{1982ApJ...262L..53F}
{Ferland}, G.~J., {Pepper}, G.~H., {Langer}, S.~H., {MacDonald}, J., {Truran},
  J.~W., \& {Shaviv}, G. 1982, \apjl, 262, L53

\bibitem[{{Frank} {et~al.}(1992){Frank}, {King}, \&
  {Raine}}]{1992apa..book.....F}
{Frank}, J., {King}, A., \& {Raine}, D. 1992, {Accretion Power in Astrophysics}
  (Cambridge: Cambridge Univ. Press)

\bibitem[{{Gaensicke} \& {Beuermann}(1996)}]{1996A&A...309L..47G}
{Gaensicke}, B.~T., \& {Beuermann}, K. 1996, \aap, 309, L47

\bibitem[{{George} \& {Fabian}(1991)}]{1991MNRAS.249..352G}
{George}, I.~M., \& {Fabian}, A.~C. 1991, \mnras, 249, 352

\bibitem[{{Green} {et~al.}(1982){Green}, {Ferguson}, {Liebert}, \&
  {Schmidt}}]{1982PASP...94..560G}
{Green}, R.~F., {Ferguson}, D.~H., {Liebert}, J., \& {Schmidt}, M. 1982, \pasp,
  94, 560

\bibitem[{{Haberl} \& {Motch}(1995)}]{1995A&A...297L..37H}
{Haberl}, F., \& {Motch}, C. 1995, \aap, 297, L37+

\bibitem[{{Haberl} {et~al.}(2002){Haberl}, {Motch}, \&
  {Zickgraf}}]{2002A&A...387..201H}
{Haberl}, F., {Motch}, C., \& {Zickgraf}, F.-J. 2002, \aap, 387, 201

\bibitem[{{Hamada} \& {Salpeter}(1961)}]{1961ApJ...134..683H}
{Hamada}, T., \& {Salpeter}, E.~E. 1961, \apj, 134, 683

\bibitem[{{Hameury} \& {King}(1988)}]{1988MNRAS.235..433H}
{Hameury}, J.~M., \& {King}, A.~R. 1988, \mnras, 235, 433

\bibitem[{{Harrison} {et~al.}(1999){Harrison}, {McNamara}, {Szkody},
  {McArthur}, {Benedict}, {Klemola}, \& {Gilliland}}]{1999ApJ...515L..93H}
{Harrison}, T.~E., {McNamara}, B.~J., {Szkody}, P., {McArthur}, B.~E.,
  {Benedict}, G.~F., {Klemola}, A.~R., \& {Gilliland}, R.~L. 1999, \apjl, 515,
  L93

\bibitem[{{Hessman} {et~al.}(1989){Hessman}, {Koester}, {Schoembs}, \&
  {Barwig}}]{1989A&A...213..167H}
{Hessman}, F.~V., {Koester}, D., {Schoembs}, R., \& {Barwig}, H. 1989, \aap,
  213, 167

\bibitem[{{Horne} {et~al.}(1994){Horne}, {Marsh}, {Cheng}, {Hubeny}, \&
  {Lanz}}]{1994ApJ...426..294H}
{Horne}, K., {Marsh}, T.~R., {Cheng}, F.~H., {Hubeny}, I., \& {Lanz}, T. 1994,
  \apj, 426, 294

\bibitem[{{Ishida} {et~al.}(1996){Ishida}, {Fujimoto}, \&
  {Matsuzaki}}]{1996cvro.coll..259I}
{Ishida}, M., {Fujimoto}, B., \& {Matsuzaki}, K. 1996, in ASSL Vol. 208,
  Cataclysmic Variables and Related Objects, ed. A.~Evans \& J.~H. Wood
  (Dordrecht: Kluwer), 259

\bibitem[{{Jansen} {et~al.}(2001)}]{2001A&A...365L...1J}
{Jansen}, F., {et~al.} 2001, \aap, 365, L1

\bibitem[{{King}(1997)}]{1997MNRAS.288L..16K}
{King}, A.~R. 1997, \mnras, 288, L16

\bibitem[{{La Dous}(1991)}]{1991A&A...252..100L}
{La Dous}, C. 1991, \aap, 252, 100

\bibitem[{{Lasota} {et~al.}(1995){Lasota}, {Hameury}, \&
  {Hure}}]{1995A&A...302L..29L}
{Lasota}, J.~P., {Hameury}, J.~M., \& {Hure}, J.~M. 1995, \aap, 302, L29

\bibitem[{{Liedahl}(1999)}]{1999xrsa.conf..189L}
{Liedahl}, D.~A. 1999, in X-Ray Spectroscopy in Astrophysics, ed. J.~van
  Paradijs \& J.~Bleeker (Springer: Amsterdam), 189

\bibitem[{{Liedahl} {et~al.}(1995){Liedahl}, {Osterheld}, \&
  {Goldstein}}]{1995ApJ...438L.115L}
{Liedahl}, D.~A., {Osterheld}, A.~L., \& {Goldstein}, W.~H. 1995, \apjl, 438,
  L115

\bibitem[{{Long} \& {Gilliland}(1999)}]{1999ApJ...511..916L}
{Long}, K.~S., \& {Gilliland}, R.~L. 1999, \apj, 511, 916

\bibitem[{{Long} {et~al.}(1994){Long}, {Sion}, {Huang}, \&
  {Szkody}}]{1994ApJ...424L..49L}
{Long}, K.~S., {Sion}, E.~M., {Huang}, M., \& {Szkody}, P. 1994, \apjl, 424,
  L49

\bibitem[{{Lynden-Bell} \& {Pringle}(1974)}]{1974MNRAS.168..603L}
{Lynden-Bell}, D., \& {Pringle}, J.~E. 1974, \mnras, 168, 603

\bibitem[{{Mahasena} \& {Osaki}(1999)}]{1999PASJ...51...45M}
{Mahasena}, P., \& {Osaki}, Y. 1999, \pasj, 51, 45

\bibitem[{{Makishima} {et~al.}(1986){Makishima}, {Maejima}, {Mitsuda}, {Bradt},
  {Remillard}, {Tuohy}, {Hoshi}, \& {Nakagawa}}]{1986ApJ...308..635M}
{Makishima}, K., {Maejima}, Y., {Mitsuda}, K., {Bradt}, H.~V., {Remillard},
  R.~A., {Tuohy}, I.~R., {Hoshi}, R., \& {Nakagawa}, M. 1986, \apj, 308, 635

\bibitem[{{Mason} {et~al.}(2001)}]{2001A&A...365L..36M}
{Mason}, K.~O., {et~al.} 2001, \aap, 365, L36

\bibitem[{{Mauche}(2002)}]{2002pcvr.conf..113M}
{Mauche}, C.~W. 2002, in ASP Conf. Ser. 261, The Physics of Cataclysmic
  Variables and Related Objects, ed. B.~T. G\"ansicke, K.~Beuermann, \&
  K.~Reinsch (San Francisco: ASP), 113

\bibitem[{{Medvedev} \& {Menou}(2002)}]{2002ApJ...565L..39M}
{Medvedev}, M.~V., \& {Menou}, K. 2002, \apjl, 565, L39

\bibitem[{{Mewe} {et~al.}(1985){Mewe}, {Gronenschild}, \& {van den
  Oord}}]{1985A&AS...62..197M}
{Mewe}, R., {Gronenschild}, E.~H.~B.~M., \& {van den Oord}, G.~H.~J. 1985,
  \aaps, 62, 197

\bibitem[{{Meyer} \& {Meyer-Hofmeister}(1994)}]{1994A&A...288..175M}
{Meyer}, F., \& {Meyer-Hofmeister}, E. 1994, \aap, 288, 175

\bibitem[{{Mitsuda} {et~al.}(1984)}]{1984PASJ...36..741M}
{Mitsuda}, K., {et~al.} 1984, \pasj, 36, 741

\bibitem[{{Mukai} {et~al.}(2003){Mukai}, {Kinkhabwala}, {Peterson}, {Kahn}, \&
  {Paerels}}]{2003ApJ...586L..77M}
{Mukai}, K., {Kinkhabwala}, A., {Peterson}, J.~R., {Kahn}, S.~M., \& {Paerels},
  F. 2003, \apjl, 586, L77

\bibitem[{{Mukai} {et~al.}(1997){Mukai}, {Wood}, {Naylor}, {Schlegel}, \&
  {Swank}}]{1997ApJ...475..812M}
{Mukai}, K., {Wood}, J.~H., {Naylor}, T., {Schlegel}, E.~M., \& {Swank}, J.~H.
  1997, \apj, 475, 812

\bibitem[{{Mushotzky} \& {Szymkowiak}(1988)}]{1988cfcg.work...53M}
{Mushotzky}, R.~F., \& {Szymkowiak}, A.~E. 1988, in Cooling flows in clusters
  and galaxies, ed. A.~C. Fabian (Dordrecht: Kluwer), 53

\bibitem[{{Narayan} \& {Popham}(1993)}]{1993Natur.362..820N}
{Narayan}, R., \& {Popham}, R. 1993, \nat, 362, 820

\bibitem[{{Pandel} {et~al.}(2003){Pandel}, {C{\' o}rdova}, \&
  {Howell}}]{2003MNRAS.346.1231P}
{Pandel}, D., {C{\' o}rdova}, F.~A., \& {Howell}, S.~B. 2003, \mnras, 346, 1231

\bibitem[{{Perna} {et~al.}(2003){Perna}, {McDowell}, {Menou}, {Raymond}, \&
  {Medvedev}}]{2003ApJ...598..545P}
{Perna}, R., {McDowell}, J., {Menou}, K., {Raymond}, J., \& {Medvedev}, M.~V.
  2003, \apj, 598, 545

\bibitem[{{Ponman} {et~al.}(1995){Ponman}, {Belloni}, {Duck}, {Verbunt},
  {Watson}, {Wheatley}, \& {Pfeffermann}}]{1995MNRAS.276..495P}
{Ponman}, T.~J., {Belloni}, T., {Duck}, S.~R., {Verbunt}, F., {Watson}, M.~G.,
  {Wheatley}, P.~J., \& {Pfeffermann}, E. 1995, \mnras, 276, 495

\bibitem[{{Porquet} \& {Dubau}(2000)}]{2000A&AS..143..495P}
{Porquet}, D., \& {Dubau}, J. 2000, \aaps, 143, 495

\bibitem[{{Ramsay} {et~al.}(2001{\natexlab{a}})}]{2001A&A...365L.288R}
{Ramsay}, G., {et~al.} 2001{\natexlab{a}}, \aap, 365, L288

\bibitem[{{Ramsay} {et~al.}(2001{\natexlab{b}})}]{2001A&A...365L.294R}
---. 2001{\natexlab{b}}, \aap, 365, L294

\bibitem[{{Ritter} \& {Kolb}(2003)}]{2003A&A...404..301R}
{Ritter}, H., \& {Kolb}, U. 2003, \aap, 404, 301

\bibitem[{{Shafter} \& {Szkody}(1984)}]{1984ApJ...276..305S}
{Shafter}, A.~W., \& {Szkody}, P. 1984, \apj, 276, 305

\bibitem[{{Sion}(1991)}]{1991AJ....102..295S}
{Sion}, E.~M. 1991, \aj, 102, 295

\bibitem[{{Sion}(1999)}]{1999PASP..111..532S}
---. 1999, \pasp, 111, 532

\bibitem[{{Sion} {et~al.}(2001){Sion}, {Cheng}, {Szkody}, {G{\" a}nsicke},
  {Sparks}, \& {Hubeny}}]{2001ApJ...561L.127S}
{Sion}, E.~M., {Cheng}, F., {Szkody}, P., {G{\" a}nsicke}, B., {Sparks}, W.~M.,
  \& {Hubeny}, I. 2001, \apjl, 561, L127

\bibitem[{{Sion} {et~al.}(1997){Sion}, {Cheng}, {Sparks}, {Szkody}, {Huang}, \&
  {Hubeny}}]{1997ApJ...480L..17S}
{Sion}, E.~M., {Cheng}, F.~H., {Sparks}, W.~M., {Szkody}, P., {Huang}, M., \&
  {Hubeny}, I. 1997, \apjl, 480, L17

\bibitem[{{Sion} {et~al.}(1994){Sion}, {Long}, {Szkody}, \&
  {Huang}}]{1994ApJ...430L..53S}
{Sion}, E.~M., {Long}, K.~S., {Szkody}, P., \& {Huang}, M. 1994, \apjl, 430,
  L53

\bibitem[{{Sproats} {et~al.}(1996){Sproats}, {Howell}, \&
  {Mason}}]{1996MNRAS.282.1211S}
{Sproats}, L.~N., {Howell}, S.~B., \& {Mason}, K.~O. 1996, \mnras, 282, 1211

\bibitem[{{Str{\"u}der} {et~al.}(2001)}]{2001A&A...365L..18S}
{Str{\"u}der}, L., {et~al.} 2001, \aap, 365, L18

\bibitem[{{Szkody} {et~al.}(2002){Szkody}, {Nishikida}, {Raymond}, {Seth},
  {Hoard}, {Long}, \& {Sion}}]{2002ApJ...574..942S}
{Szkody}, P., {Nishikida}, K., {Raymond}, J.~C., {Seth}, A., {Hoard}, D.~W.,
  {Long}, K.~S., \& {Sion}, E.~M. 2002, \apj, 574, 942

\bibitem[{{Thorstensen}(1986)}]{1986AJ.....91..940T}
{Thorstensen}, J.~R. 1986, \aj, 91, 940

\bibitem[{{Turner} {et~al.}(2001)}]{2001A&A...365L..27T}
{Turner}, M.~J.~L., {et~al.} 2001, \aap, 365, L27

\bibitem[{{Tylenda}(1981)}]{1981AcA....31..127T}
{Tylenda}, R. 1981, Acta Astronomica, 31, 127

\bibitem[{{van Teeseling} {et~al.}(1996){van Teeseling}, {Beuermann}, \&
  {Verbunt}}]{1996A&A...315..467V}
{van Teeseling}, A., {Beuermann}, K., \& {Verbunt}, F. 1996, \aap, 315, 467

\bibitem[{{van Teeseling} \& {Verbunt}(1994)}]{1994A&A...292..519V}
{van Teeseling}, A., \& {Verbunt}, F. 1994, \aap, 292, 519

\bibitem[{{Verbunt}(1987)}]{1987A&AS...71..339V}
{Verbunt}, F. 1987, \aaps, 71, 339

\bibitem[{{Wade}(1982)}]{1982AJ.....87.1558W}
{Wade}, R.~A. 1982, \aj, 87, 1558

\bibitem[{{Warner}(1987)}]{1987MNRAS.227...23W}
{Warner}, B. 1987, \mnras, 227, 23

\bibitem[{{Warner}(1995)}]{1995cvs..book.....W}
---. 1995, {Cataclysmic variable stars} (Cambridge: Cambridge Univ. Press)

\bibitem[{{Warner} \& {Wickramasinghe}(1991)}]{1991MNRAS.248..370W}
{Warner}, B., \& {Wickramasinghe}, D.~T. 1991, \mnras, 248, 370

\bibitem[{{Wheatley} \& {West}(2003)}]{2003MNRAS.345.1009W}
{Wheatley}, P.~J., \& {West}, R.~G. 2003, \mnras, 345, 1009

\bibitem[{{Wu}(2000)}]{2000SSRv...93..611W}
{Wu}, K. 2000, Space Science Reviews, 93, 611

\end{thebibliography}

\end{document}